\documentclass[longauth]{aa}  
\usepackage{graphicx}
\usepackage{epstopdf}
\usepackage{subfigure}
\usepackage{enumitem}
\usepackage{txfonts}
\usepackage{booktabs}
\usepackage{graphics}
\usepackage{sidecap}

\graphicspath{{./images/}}
\DeclareGraphicsExtensions{.eps}

\def\ltsima{$\; \buildrel < \over \sim \;$}
\def\simlt{\lower.5ex\hbox{\ltsima}}
\def\gtsima{$\; \buildrel > \over \sim \;$}
\def\simgt{\lower.5ex\hbox{\gtsima}}

\def\ergs{{erg s$^{-1}$}}
\def\cm2{{cm$^{-2}$}}

\def\lsun{{$L_{\rm \odot}$}}
\def\lbol{{$L_{\rm Bol}$}}

\def\kms{{km s$^{-1}$}} 
\def\msun{{$M_{\rm \odot}$}}
\def\55{J1555$+$1003}
\def\38{J1538$+$0855}
\def\mbh{$M_{\rm BH}$}
\def\43{CO(4$-$3)}

\def\mum{{$\mu$m}}
\def\mjb{{mJy beam$^{-1}$}}
\def\cofive{{CO(5$-$4)}}
\def\cofour{{CO(4$-$3)}}
\def\coone{{CO(1$-$0)}}
\def\cii{{[CII]}}
\def\jzdue{{J0209$-$0005}}
\def\jzotto{{J0801$+$5210}}
\def\jdsette{{J1701$+$6412}}
\def\jquatt{{J1433$+$0227}}
\def\jdieci{{J1015$+$0020}}
\def\jqcinque{{J1555$+$1003}}
\def\jqnove{{J1549$+$1245}}
\def\jqotto{{J1538$+$0855}}
\def\jsedici{{J1639$+$2824}}
\def\kms{{km~s$^{-1}$}}
\def\msunyr{{$M_\odot$ yr$^{-1}$}}
\def\mgas{{$M_{\rm gas}$}}
\begin{document}

  \title{The WISSH QSOs project}

   \subtitle{IX. Cold gas content and environment of luminous QSOs at $z\sim2.4-4.7$}

\author{M. Bischetti \inst{1,2} \thanks{e-mail: manuela.bischetti@inaf.it}
         \and C. Feruglio   \inst{1}
         \and E. Piconcelli \inst{2}
         \and F. Duras      \inst{3,2}
         \and M. P\'erez-Torres  \inst{4,5}
         \and R. Herrero    \inst{6,7}
         \and G. Venturi    \inst{8}
         \and S. Carniani   \inst{9}
         \and G. Bruni      \inst{10}
         \and I. Gavignaud  \inst{11}
         \and V. Testa      \inst{2}
         \and A. Bongiorno  \inst{2}
         \and M. Brusa      \inst{12,13}
         \and C. Circosta   \inst{14}
         \and G. Cresci     \inst{15}
         \and V. D'Odorico  \inst{1,9}
         \and R. Maiolino   \inst{16,17}
         \and A. Marconi    \inst{18,15}
         \and M. Mingozzi   \inst{19}
         \and C. Pappalardo \inst{20}
         \and M. Perna      \inst{21,15}
         \and E. Traianou   \inst{22}
         \and A. Travascio  \inst{2}
         \and G. Vietri     \inst{23}
         \and L. Zappacosta \inst{2}
         \and F. Fiore      \inst{1}
         }

   \institute{INAF - Osservatorio Astronomico di Trieste, Via G. B. Tiepolo 11, I--34143 Trieste, Italy
   \and INAF - Osservatorio Astronomico di Roma, Via Frascati 33, I--00078 Monte Porzio Catone, Italy 
   \and Dipartimento di Matematica e Fisica, Universit\'a Roma Tre, Via della Vasca Navale 84, 00146, Roma, Italy 
   \and Instituto de Astrofísica de Andaluc\'ia (IAA, CSIC), Glorieta de las Astronom\'ia, s/n, E-18008 Granada, Spain
   \and Departamento de F\'isica Teorica, Facultad de Ciencias, Universidad de Zaragoza, Spain
   \and European Southern Observatory (ESO), Alonso de C\'ordova 3107, Vitacura, Casilla 19001, Santiago de Chile, Chile \and Institute of Space Sciences (ICE, CSIC), Campus UAB, Carrer de Magrans, E-08193 Barcelona, Spain
   \and Instituto de Astrof\'isica, Pontificia Universidad Cat\'olica de Chile, Avda Vicuna Mackenna 4860, 8970117 Macul, Santiago, Chile
   \and Scuola Normale Superiore, Piazza dei Cavalieri 7, I-56126 Pisa, Italy
   \and INAF - Istituto di Astrofisica e Planetologia Spaziali, Via Fosso del Cavaliere 100, 00133 Roma, Italy
   \and Dpto. de Ciencias Fisicas, Universidad Andres Bello, Campus La Casona, Fernandez Concha 700, 7500912 Santiago, Chile
   \and Dipartimento di Fisica e Astronomia, Alma Mater Studiorum Universit\'a di Bologna,  via Gobetti 93/2,  40129 Bologna, Italy 
   \and INAF - Osservatorio di Astrofisica e Scienza dello Spazio di Bologna, via Gobetti 93/3,  40129 Bologna, Italy 
   \and Department of Physics \& Astronomy, University College London, Gower Street, London WC1E 6BT, United Kingdom
   \and INAF-Osservatorio Astrofisico di Arcetri, Largo E. Fermi 2, 50125, Firenze, Italy
   \and Kavli Institute for Cosmology, University of Cambridge, Madingley Road, Cambridge, CB3 0HA, UK
   \and Cavendish Laboratory, University of Cambridge, 19 J. J. Thomson Avenue, Cambridge CB3 0HE, UK
   \and Dipartimento di Fisica e Astronomia, Universit\'a degli Studi di Firenze, Via G. Sansone 1, 50019, Sesto Fiorentino, Firenze, Italy
   \and INAF - Padova Astronomical Observatory, Vicolo dell’Osservatorio 5, I-35122 Padova, Italy
   \and Instituto de Astrofísica e Ci\^encias do Espa\c{o}, Universidade de Lisboa, OAL, Tapada da Ajuda, 1349-018, Lisboa, Portugal
   \and Centro de Astrobiolog\'ia (CAB, CSIC-INTA), Dpto. de Astrof\'\i sica, Ctra de Ajalvir km~4, 28850, Torrej\'on de Ardoz, Madrid, Spain 
   \and Max-Planck-Institut f\"ur Radioastronomie, Auf dem H\"ugel 69, D-53121 Bonn, Germany
   \and INAF - Istituto di Astrofisica Spaziale e Fisica cosmica Milano, via A. Corti 12, 20133, Milano, Italy
   }

   \date{}

 
  \abstract
  {Sources at the brightest end of the quasi-stellar object (QSO) luminosity function, during the peak epoch in the history of star formation and black hole accretion ($z\sim2-4$, often referred to as "Cosmic noon") are privileged sites to study  the cycle of feeding \& feedback processes in massive galaxies.}
  {We aim to perform the first systematic study of cold gas properties in the most luminous QSOs, by characterising their host-galaxies and environment. These targets exhibit indeed widespread evidence of outflows at nuclear and galactic scales.}
  {We analyse ALMA, NOEMA and JVLA observations of the far-infrared continuum, CO and [CII] emission lines in eight QSOs (bolometric luminosity $L_{\rm Bol}\gtrsim3\times10^{47}$ \ergs) from the WISE-SDSS selected hyper-luminous (WISSH) QSOs sample at $z\sim2.4-4.7$.}
  {We report a $100\%$ emission line detection rate and a $80\%$ detection rate in continuum emission, and we find CO emission to be consistent with the steepest CO ladders observed so far. Sub-millimetre data reveal presence of (one or more) bright companion galaxies around $\sim80\%$ of WISSH QSOs, at projected distances of $\sim6-130$ kpc. We observe a variety of sizes for the molecular gas reservoirs ($\sim1.7-10$ kpc), mostly associated with rotating disks with disturbed kinematics.
  WISSH QSOs typically show lower CO luminosity and higher star formation efficiency than infrared matched, $z\sim0-3$ main-sequence galaxies, implying that, given the observed SFR $\sim170-1100$ \msunyr, molecular gas is converted into stars in $\lesssim50$ Myr. Most targets show extreme dynamical to black-hole mass ratios $M_{\rm dyn}/M_{\rm BH}\sim3-10$, two orders of magnitude smaller than local relations. 
  The molecular gas fraction in the host-galaxies of WISSH is lower by a factor of $\sim10-100$ than in star forming galaxies with similar $M_{*}$. }
  {Our analysis reveals that hyper-luminous QSOs at Cosmic noon undergo an intense growth phase of both the central super-massive black hole and of the host-galaxy. These systems pinpoint the high-density sites where giant galaxies assemble, where we show that mergers play a major role in the build-up of the final host-galaxy mass. We suggest that the observed low molecular gas fraction and short depletion timescale are due to AGN feedback, whose presence is indicated by fast AGN-driven ionised outflows in all our targets. }

  \keywords{galaxies:~high-redshift -- galaxies:~ISM -- quasars:~emission lines -- quasars:~supermassive black holes -- submillimeter: galaxies -- techniques:~interferometric}
 \maketitle
%

\section{Introduction}


Our understanding of the complex interplay between super-massive black hole (SMBH) activity, properties of the host-galaxy interstellar medium (ISM) and of the circum-galactic environment is still incomplete. According to models of active galactic nucleus (AGN) and galaxy co-evolution, two main processes are expected to drive the evolution of massive galaxies (stellar mass $M_*>10^{11}$ \msun): (i) galaxy interactions and mergers, and (ii) radiative and mechanical feedback related to AGN activity \citep[e.g.][]{Croton06, Sijacki07,Martin-Navarro18,Choi18}, during the growth of heavy SMBHs \citep[$M_{\rm BH}/M_*\sim10^{-3}$, where $M_{\rm BH}$ is the black hole mass, e.g.][]{Haring04}. These two processes are tightly correlated, as galaxy interactions destabilise the gas and make it available fuel for both star formation (SF) and nuclear accretion \citep[e.g.][]{Volonteri15, AnglesAlcazar17}. In turn, powerful AGN-driven, galaxy-scale outflows can hamper further SF and nuclear gas accretion by injecting energy and entropy into the host-galaxy ISM \citep[e.g.][]{King15,Richings18,Menci19} and circum-galactic environment \citep{vandeVoort17,Travascio20}.

It is crucial to probe the cycle of feeding \& feedback processes in massive galaxies and the AGN-galaxy co-evolution during Cosmic noon ($z\sim2-4.5$), that is the peak epoch of galaxy assembly and SMBH accretion. Indeed, hyper-luminous quasi-stellar object (QSOs, bolometric luminosity $L_{\rm Bol}>10^{47}$ \ergs) are preferred targets, being powered by massive ($M_{\rm BH}\gtrsim10^9$ \msun), highly accreting SMBHs \citep{Banerji15,Shen16,Vietri18}. These heavy SMBHs have been found to reside in most local giant elliptical galaxies \citep{Kormendy13,Shankar19}, although most of their mass has been assembled at early times during bright AGN phases \citep[e.g.][]{Marconi04,Delvecchio14}. Giant ellipticals have also built most of their stellar mass at $z\gtrsim2$ ($\sim80$ \% for sources with $M_*>10^{11}$ \msun) and then rapidly evolved off the galaxy Main Sequence (MS) becoming “red \& dead” \citep[e.g.][]{Santini09,McDermid15}. Overall, this strongly suggests that high-$z$, hyper-luminous QSOs are the assembly sites of giant galaxies. Indeed, the huge luminosities of these QSOs are likely triggered by galaxy interactions \citep[e.g.][]{Urrutia08,Menci14}, and drive powerful outflows which may affect the host-galaxy \citep{Carniani15,Zakamska16,Bischetti17,Bischetti19,Perrotta19,Villar-Martin20,Herrera-Camus20}.

Investigating the properties of the cold ISM (that is the raw material of SF) in objects at the brightest end of the QSO luminosity function is mandatory to constrain the impact of AGN activity on the host-galaxy evolution. Our knowledge of the host-galaxies of $z>1$ QSOs has been revolutionised by interferometric observations at sub-millimetre wavelengths. Specifically, CO rotational emission is the most targeted transition in $z\lesssim4$ QSOs and probes the molecular gas reservoir. At higher redshift, it is possible to more easily exploit the bright [CII] fine structure emission line at $\lambda158$ $\mu$m, which is mostly associated with cold neutral gas \citep{Solomon05, Carilli&Walter13,Zanella18,Veilleux20}. 
Previous studies of luminous QSOs at $z\sim2-6$ revealed a variety of host-galaxy properties. These sources typically reside in compact hosts, in some of which rotating disks are in place, while others show irregular morphology and disturbed kinematics \citep{Diaz-Santos16,Banerji17,Brusa18,Feruglio18,Talia18,Pensabene20}. Intense SF activity of hundreds to thousands solar masses per year is usually observed \citep{Maiolino12,Duras17,Fan19,Nguyen20}, although it is unclear whether this is linked to galaxy interactions \citep{Trakhtenbrot17,Bischetti18,Fan18,Fan19}.

While great emphasis has been put in uncovering the formation and early growth of QSOs and their host-galaxies at $5\lesssim z\lesssim 7$ \citep[ e.g.][]{Wang13,Wang16,Decarli18,Venemans17,Venemans19,Banados19}, currently there is a lack of systematic investigation into the ISM and environment properties of QSOs shining at Cosmic noon. There is a series of works based on individual, often gravitationally-lensed sources \citep[e.g.][and references therein]{Carilli&Walter13} or peculiar AGN types, such as heavily-reddened and hot, dust-obscured QSOs \citep[e.g.][]{Banerji17,Fan18}. Few studies, based on heterogeneous samples including QSOs with different selection criteria \citep[][]{Feruglio14,Perna18} suggested a typically higher star formation efficiency (SFE) in QSOs, defined as the rate of SF per unit of molecular gas mass, with respect to non-active sources on the galaxy MS, although e.g. \cite{Kirkpatrick19} found no significant correlation between the SFE and the AGN fraction at mid-infrared wavelengths. An increased SFE in QSOs is expected in the standard paradigm of a QSO expelling/heating gas from the host-galaxy in the transitory phase from starburst to blue QSO \citep{DiMatteo07,Hopkins12}. Indeed,
few works probing the molecular gas content in $z\sim1.5-2.5$ QSOs with known presence of AGN-driven outflows reported a lower molecular gas content with respect to MS galaxies \citep{Brusa15,Brusa18,Carniani17,Kakkad17}, while others found no difference \citep[e.g.][]{Herrera-Camus19}.

In this framework, we have undertaken the WISE-SDSS selected hyper-luminous (WISSH) QSOs project to study the most powerful AGN in the Universe, which happen to shine at Cosmic noon and show widespread evidence of strong, AGN-driven outflows on nuclear to circum-galactic scales \citep[][]{Bischetti17,Vietri18,Bruni19,Travascio20}. These QSOs are therefore ideal targets to shed light on the AGN-galaxy feeding and feedback cycle. In this work, we present sub-millimetre and millimetre observations of eight WISSH QSOs at $z\sim2.4-4.7$ to probe SMBH and galaxy assembly at the massive end of the mass function and investigate the interplay between nuclear activity and the physical properties of the host-galaxy ISM.

The paper is organised as follows. In Sect. \ref{sect:sample} we describe the targets and the (sub-)millimetre observations, detailing data reduction and analysis in Sect. \ref{sect:data}. The main results are presented in Sect. \ref{sect:results}, focusing on continuum (\ref{sect:contmaps}), CO and [CII] emission (\ref{sect:linemaps}). The CO spectral line energy distribution (SLED) and the UV-to-FIR spectral energy distribution (SED) are described in Sect. \ref{sect:sled} and Sect. \ref{sect:sed}. Sect. \ref{sect:kine} is dedicated to the cold gas kinematics. We discuss environment, star-formation efficiency, gas excitation, dynamical mass and molecular gas fraction of WISSH QSOs in Sect. \ref{sect:discussion}. Finally, our conclusions are summarised in Sect. \ref{sect:conclusion}.

Throughout this paper, we assume a $\rm\Lambda CDM$ cosmology with $H_0 = 67.3$ \kms\ Mpc$^{-1}$, $\Omega_{\Lambda} = 0.69$ and $\Omega_{\rm M}= 0.31$ \citep{Planck16}.

\section{Sample and Observations}\label{sect:sample}

\begin{table*}[htb]
\centering
\caption{Journal of observations. Columns give the following information: (1) SDSS ID, (2-3) celestial coordinates, (4) redshift based on SDSS DR12 \citep{Alam15}, (5) observation telescope, (6) observed transition, (7) angular resolution, (8) observed continuum frequency, (9) rms sensitivity for a channel width of 40 \kms, and (10) continuum rms sensitivity.}
\setlength{\tabcolsep}{4 pt}

\begin{tabular}{lccccccccc}
\toprule
ID & RA & Dec & $z_{\rm SDSS}$ &  Telescope & Transition & Beamsize & $\nu_{\rm cont}$ & $\sigma_{\rm 40\ km\ s^{-1}}$ & $\sigma_{\rm cont}$\\
 & & & &  & & arcsec$^2$ & GHz & {\small \mjb} & {\small \mjb} \\
 (1) & (2) & (3) & (4) & (5) & (6) & (7) & (8) & (9) & (10) \\
 
\midrule
J0209$-$0005  & 02:09:50.71 & $-$00:05:06.22 & 2.849 &  NOEMA  & \cofive & $4.9\times2.7$   & 148.7 & 0.90 & 0.040 \\
              &             &                &       &  JVLA   & \coone  & $3.1\times2.3$   & 31.6  & 0.08 & 0.005 \\   
J0801$+$5210  & 08:01:17.82 & $+$52:10:34.94 & 3.217 &  NOEMA  & \cofive & $3.8\times1.9$   & 136.6 & 1.05 & 0.100 \\
              &             &                &       &  JVLA   & \coone  & $3.4\times2.3$   & 30.2  & 0.16 & 0.010 \\ 
J1433$+$0227  & 14:33:52.21 & $+$02:27:14.01 & 4.622 &  ALMA   & \cii    & $0.44\times0.34$ & 338.1 & 0.43 & 0.051 \\
J1538$+$0855  & 15:38:30.49 & $+$08:55:17.42 & 3.542 &  ALMA   & \cofour & $1.1\times0.7$   & 94.3  & 0.49 & 0.028 \\
J1549$+$1245  & 15:49:38.71 & $+$12:45:09.25 & 2.386 &  ALMA   & CO(4$-$3) & $1.0\times0.9$ & 141.8 & 0.10 & 0.007 \\
J1555$+$1003  & 15:55:14.86 & $+$10:03:51.23 & 3.512 &  ALMA   & \cofour & $1.0\times0.7$   & 94.30 & 0.57 & 0.028 \\
J1639$+$2824  & 16:39:09.11 & $+$28:24:47.16 & 3.786 &  ALMA   & \cofour & $0.20\times0.13$ & 100.4 & 0.43 & 0.026 \\
J1701$+$6412  & 17:01:00.60 & $+$64:12:09.32 & 2.724 &  NOEMA  & \cofive & $4.0\times2.1$   & 154.7 & 1.03 & 0.100 \\
J1015$+$0020$^*$ & 10:15:49.00 & $+$00:20:20.03 & 4.400 & ALMA &  \cii   & $0.22\times0.18$ & 357.3 & 0.19 & 0.040 \\
\bottomrule
\end{tabular}
\flushleft 
\footnotesize $^*$ presented in \cite{Bischetti18}
\label{table:sample}
\end{table*}

\begin{figure}[htb]
    \centering
    \includegraphics[width =1\columnwidth]{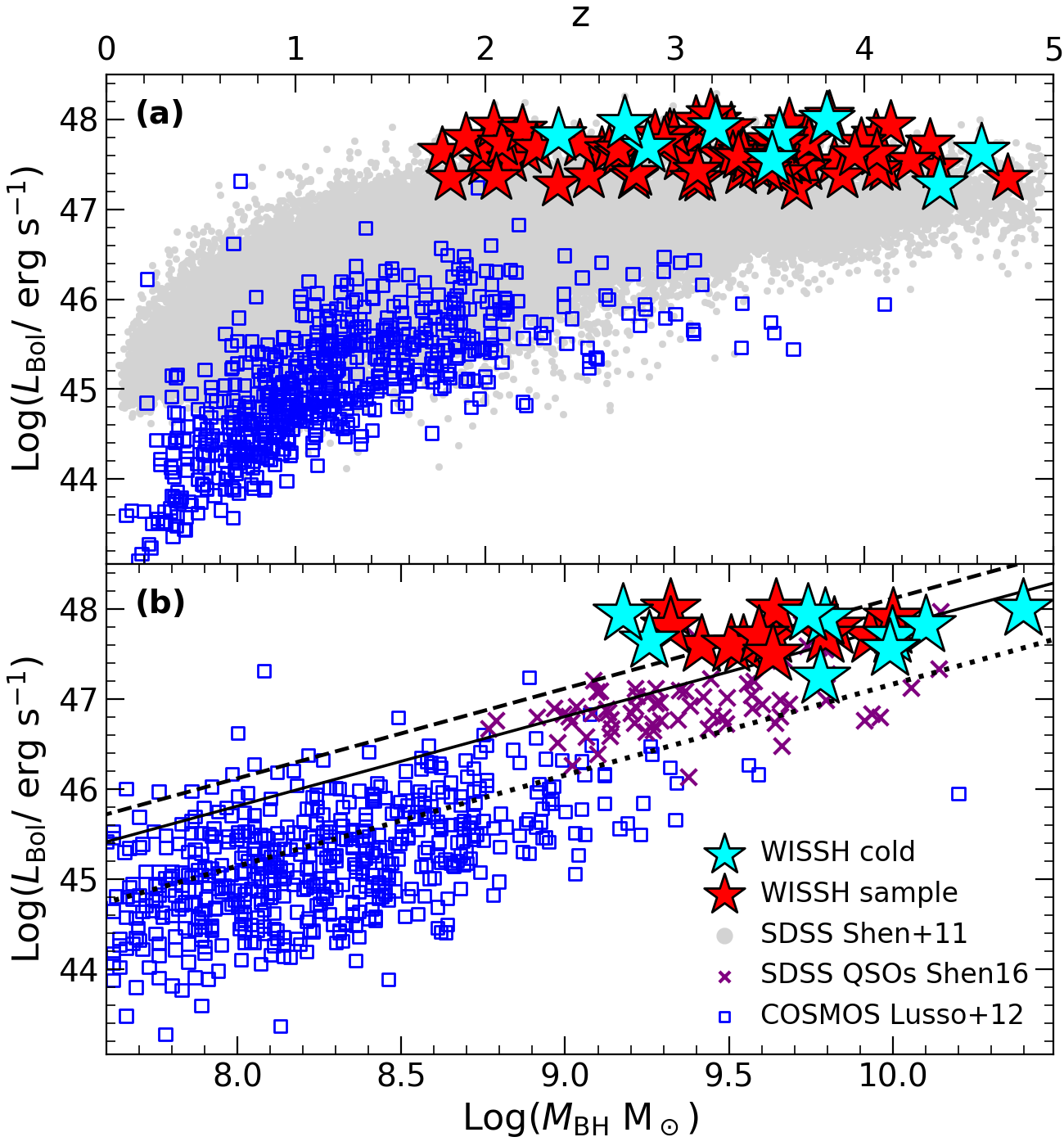}
    \caption{{\bf Panel (a):} bolometric luminosity as a function of redshift for the total WISSH sample, compared to SDSS QSOs from \cite{Shen11} and COSMOS QSOs from \cite{Lusso12}. Cyan stars indicate WISSH QSOs with (sub-)millimetre observations. Panel (b): bolometric luminosity vs black hole mass. We include WISSH QSOs with H$\beta$-based $M_{\rm BH}$ from \cite{Vietri18} and luminous SDSS QSOs from \cite{Shen16}. Lines correspond to $L_{\rm Bol}/L_{\rm Edd} = 1$ (dashed), $L_{\rm Bol}/L_{\rm Edd} = 0.5$ (solid), and $L_{\rm Bol}/L_{\rm Edd} = 0.1$ (dotted).}
    \label{fig:lbolz}
\end{figure}

\subsection{{\bf Sample description}}
The targets presented in this work are drawn from the WISSH QSOs sample \citep{Bischetti17}, including 86 hyper-luminous, Type I QSOs at $z\sim1.8-4.8$, selected to have a flux density S$_{\rm 22 \mu m}>3$ mJy by cross correlating the WISE all sky source catalogue with the SDSS DR7 catalogue. As shown in Fig. \ref{fig:lbolz}a, the WISSH sample collects the most luminous AGN known, characterised by  bolometric luminosities \lbol\ $>10^{47}$ erg~s$^{-1}$ \citep{Duras17}. The selection of Type 1 SDSS QSOs includes sources affected by low-moderate extinction \citep{Duras17}, for a large fraction of which we collected reliable estimates of the SMBH mass via broad emission line-widths. Near-IR spectroscopy revealed that WISSH QSOs are powered by highly accreting  SMBHs (accretion rate $\lambda_{\rm Edd}=L_{\rm Bol}/L_{\rm Edd}\simeq0.4-3$  (Fig. \ref{fig:lbolz}b), where $L_{\rm Edd}$ is the Eddington luminosity) with typical masses \mbh$>$2$\times10^9$ \msun\ \citep{Bischetti17, Bischetti18, Vietri18}.

Here we present observations of WISSH QSOs performed with the Atacama Large Millimetre/submillimetre Array (ALMA), the Northern Emisphere Millimetre Array (NOEMA), and the Karl G. Jansky Very Large Array (JVLA). We include ALMA band 3 data from projects 2013.1.00417.S (P.I. I. Gavignaud) and 2019.1.01070.S (P.I. G. Venturi), and band 7 data from project 2016.1.00718.S (P.I. F. Fiore); NOEMA band 2 data from projects S17BW and W17DT (P.I. G. Bruni) and JVLA Ka band data from project VLA/18A$-$028 (P.I. M. Bischetti). We also collected all publicly available archival data from the ALMA archive as of December 2019, including ALMA projects 2015.1.01602.S (band 3, P.I. M. Schramm), 2016.1.01515.S (band 7, P.I. P. Lira), and ACA project 2018.1.01806.S (band 6, P.I. K. Hall).

The combination of these projects has provided us with observations of sub-millimetre to centimetre continuum (observed frequency range $\sim350-30$ GHz) and cold gas emission for a sample of nine WISSH QSOs (Fig. \ref{fig:lbolz}). Specifically, CO rotational emission 
is available for seven targets at $z\sim2.7-3.8$, including \cofour($\nu_{\rm rest}=461.04$ GHz) and  \cofive($\nu_{\rm rest} = 576.27$ GHz) transitions, as listed in Table \ref{table:sample}. For two QSOs, namely \jzdue\ and \jzotto, we have combined information about the ground transition \coone($\nu_{\rm rest} = 115.27$ GHz) and CO($5-4$). Moreover, we have collected observations of the \cii($\nu_{\rm rest} = 1900.54$ GHz) fine structure emission line for two WISSH QSOs at $z>4$. The results about J1015$+$0020 at $z\sim4.4$ were presented in \cite{Bischetti18}, while in this work we analyse \jquatt\ at $z\sim4.7$ \citep[see also][]{Nguyen20}.

The sub-sample of WISSH QSOs analysed here derives from different projects, originally designed with different purposes. The three targets \jqotto, \jqnove\ and \jqcinque\ were selected to have AGN-driven outflows in the ionised gas phase \citep{Bischetti17,Vietri18}. Our analysis revealed that all WISSH QSOs in which we investigated the presence of [OIII] or CIV outflows satisfy this criterion \citep{Vietri18}. These three QSOs (out of nine targets) also show broad absorption line (BAL) features in their UV spectra, consistently with the BAL fraction of $\sim25\%$ measured for the parent WISSH sample by \cite{Bruni19}.
 Targets \jzdue, \jzotto, \jquatt, \jdsette\ \citep[and \jdieci\ in][]{Bischetti18} were chosen to be detected by {\it Herschel}/SPIRE photometry in the far-infrared. Nevertheless, $\sim90$\% of WISSH QSOs with {\it Herschel}/SPIRE coverage are detected \citep{Duras17}. The QSO \jsedici\ was selected to have a nearby bright star, suitable for adaptive optics (AO)-assisted observations, in order to perform a joint Subaru and ALMA study of the host galaxy properties and of the SMBH-galaxy mass ratio \citep{Schramm19}. Fig. \ref{fig:lbolz}a shows that our targets well represent the $z$ and $L_{\rm Bol}$ range covered by the total WISSH sample. Similarly, their $M_{\rm BH}$ span over a similar interval than that covered by WISSH QSOs with H$\beta$-based $M_{\rm BH}$ \citep{Vietri18}, except for the more massive SMBH in \jsedici\ (Fig. \ref{fig:lbolz}b). We note that the $M_{\rm BH}$ of \jzotto, \jqotto\ and \jqnove\ is derived from the H$\beta$ emission line \citep{Vietri18}, that of \jquatt\ is MgII-based \citep{Trakhtenbrot11}, while the $M_{\rm BH}$ of the remaining targets is CIV-based \citep{Weedman12} and corrected for the presence of CIV outflows \citep{Coatman17}. Globally, the selection of our targets does not introduce strong biases and the results of the following analysis can be considered as representative of the total WISSH sample.
 
\subsection{Data reduction and analysis}\label{sect:data}

ALMA data were calibrated using the CASA software \citep{McMullin07} in the pipeline mode. For each source, we used the CASA version indicated by the ALMA observatory and the default phase, bandpass and flux calibrators. The absolute flux accuracy is better than 10\%.
To estimate the far-IR continuum emission, we averaged visibilities over all spectral windows excluding the spectral range covered by CO or \cii\ emission lines, for a total of $\sim7$ GHz.
Moreover, to model the continuum emission next to the line, we combined the adjacent spectral windows in the ALMA baseband containing CO or \cii. We performed a fit in the uv plane to all channels with velocity $|v|>500$ \kms\ by adopting a first order polynomial continuum model. We subtracted this fit to produce continuum-subtracted visibilities.

Continuum-subtracted data cubes were created with the CASA 5.4.0 task {\it tclean}, using the {\it hogbom} cleaning algorithm \citep{Hogbom74} in non-interactive mode, and a threshold equal to three times the rms sensitivity. For all sources except \jsedici, a natural weighting of the visibilities and a common spectral channel width of 40 \kms\ were chosen. The same deconvolution procedure was adopted to produce continuum maps.
In the case of \jsedici, we applied a tapering of the visibilities beyond 900 k$\lambda$, corresponding to a baseline length $\gtrsim2800$ km, to decrease the very high native resolution ($0.11\times0.06$ arcsec$^2$) to a tapered beam of $0.20\times0.13$ arcsec$^2$. This resulted into a higher significance of CO(4$-3$) detection by a factor of $\sim1.5$, similarly to what found by \cite{Schramm19} for this QSO.

Regarding NOEMA data, project S17BW was acquired with the WideX correlator and span a total spectral range of $\sim4$ GHz, while project W17DT was obtained with the PolyFiX correlator and span a wider spectral interval of $\sim14$ GHz. We calibrated the visibilities using the CLIC pipeline of the GILDAS software (www.iram.fr/IRAMFR/GILDAS). The absolute flux accuracy is better than 10\%. Imaging was performed with MAPPING, following the same procedure described above for ALMA observations. 

Given the redshift of \jzdue\ and \jzotto, we used JVLA Ka-band observations. We observed each QSO using two separate windows of eight 128 MHz wide sub-bands for a total bandwidth of 1 GHz per window.  Each sub-band had a native spectral resolution of 1 MHz. We calibrated the visibilities, both continuum and spectral line CO(1$-$0), following standard procedures within the CASA software package. The absolute flux accuracy is better than 15\%. We imaged our sources using the  CASA task tclean, following the same procedures applied to our ALMA data.
Table \ref{table:sample} presents the journal of observations. For each target, we list the resulting beamsize, the rms sensitivity of the datacube for a channel width of 40 \kms\ ($\sigma_{\rm 40\ km s^{-1}}$) and the rms sensitivity of the continuum map ($\sigma_{\rm cont}$).

To derive flux density and size of the continuum emission in our targets, we (i) fitted the data in the image plane with a two-dimensional Gaussian profile  by using CASA task {\it imfit}, and (ii) performed a fit of the visibilities in the uv plane  with MAPPING task uvfit. In (ii), channels associated with the continuum emission, were spectrally and temporally averaged (to 30 s). We then fitted point source or Gaussian source models to the visibilities.  In the case of multiple sources, we first fitted and subtracted emission from the QSO host-galaxy, then fitted emission from nearby emitters. A Gaussian model was preferred when resulting into an increased significance of the detection and a size different from null size at $>3\sigma$ (where $\sigma$ is the error associated with the Gaussian axis derived from the fit in the uv plane). We checked that subtracting the models resulted into uniform residual noise maps. Similarly, we measured the integrated CO or [CII] flux density by averaging channels related to line emission  (filled histogram in Fig. \ref{fig:spectra}) and fitting the data in both image and uv plane.
We verified that (i) and (ii) give consistent sizes and flux density values. 

One-dimensional spectra were extracted from the continuum-subtracted datacubes from a region centred on the peak of CO or [CII] integrated flux density, including emission detected at $\rm SNR>2$. Given the modest signal-to-noise ratio of most spectra (see Fig. \ref{fig:spectra}), we fitted them with a Gaussian profile by using the python package scipy.optimize.curve\_fit leaving all parameters free to vary.   We verified that by fitting the spectra with a two-Gaussian model, the line significance and parameters were consistent within the uncertainties with the single Gaussian fit. This provided us with FWHM and centroid of the emission line profiles, which was used to infer the CO or \cii\ based redshift ($z_{\rm cold}$) of our targets listed in Table \ref{table:co-properties}.


\section{Results from sub-mm and mm observations}\label{sect:results}

\begin{figure*}[htb]
    \centering
    \includegraphics[width=0.8\textwidth]{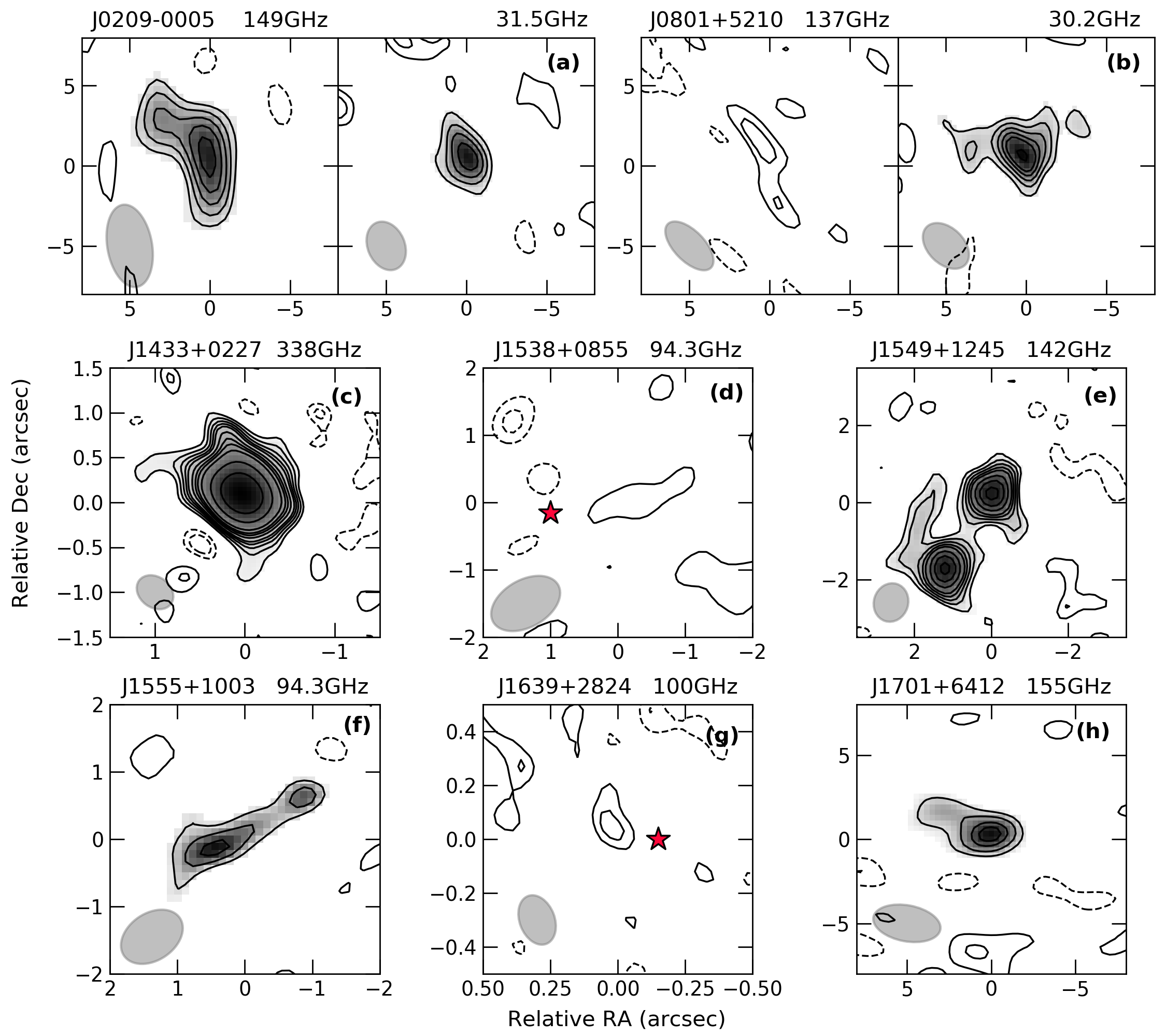}
    \caption{Maps of the continuum emission for the WISSH QSOs analysed in this work. For each panel, the observed continuum frequency is indicated by the top label. Grayscale defines the region where emission from the QSO host-galaxy and nearby continuum emitters is detected with SNR $\gtrsim4.5$. Black contours correspond to [-3, -2, 2, 3, 4, 5, 6, 8, 12, 16, 32,...]$\sigma_{\rm cont}$, where $\sigma_{\rm cont}$ values are listed in Table \ref{table:sample}. Dashed contours are for negative values. The beam of each observation is also shown by the grey ellipse. In panels (d) and (g), the red star indicates the QSO optical position from SDSS.}
    \label{fig:contmaps}
\end{figure*}

\begin{figure*}[htb]
    \centering
    \includegraphics[width=0.8\textwidth]{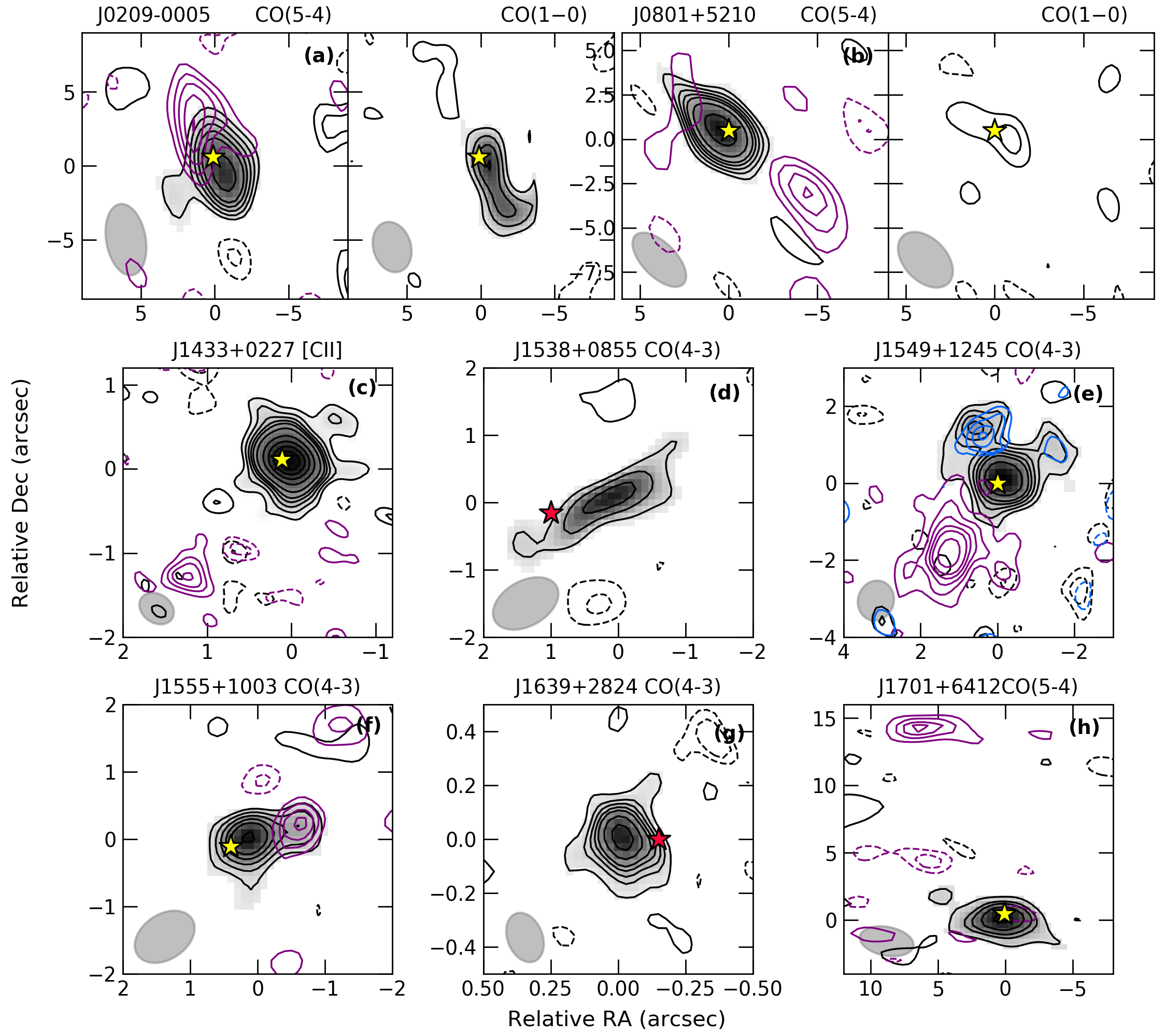}
    \caption{Maps of the cold gas emission for the WISSH QSOs analysed in this work, obtained by integrating over the spectral range covered by CO or [CII] emission. For each source, the targeted transition is indicated by the top label. Grayscale defines the region where line emission from the QSO host-galaxy is detected with SNR $\gtrsim4.5$.
    Black contours correspond to [-3, -2, 2, 3, 4, 5, 6, 8, 12, 16, 32, ...]$\sigma_{\rm cold}$ in the QSO map, where $\sigma_{\rm cold}=0.16,0.012$ Jy beam$^{-1}$ \kms\ for CO(5-4) and CO(1-0) observations in panel (a) and  $0.22,0.032$ Jy beam$^{-1}$ \kms\ in panel (b).  In panels (c-h) $\sigma_{\rm cold}= 0.091,0.098,0.021,0.11,0.082,0.18$ Jy beam$^{-1}$ \kms, respectively. Similarly, magenta and blue contours identify emission from companion line emitters. In panel (e) blue contours have been created by masking QSO emission in a circular aperture of 1 arcsec radius centred on the QSO location.
    Dashed contours are for negative values. Yellow(red) stars indicate the QSO position as traced by the millimetre(optical) continuum. The beam of each observation is also shown by the grey ellipse.}
    \label{fig:linemaps}
\end{figure*}

\subsection{Continuum emission}\label{sect:contmaps}

Figure \ref{fig:contmaps} shows the (sub-)millimetre continuum maps for the eight WISSH QSOs analysed in this work. Observed $\sim90-340$ GHz (rest-frame $\sim430-1935$ GHz) continuum emission is detected at $>4\sigma_{\rm cont}$ significance in five out of eight sources (namely \jzdue, \jquatt, \jqnove, \jqcinque, and \jdsette) while emission is observed at $\sim3\sigma_{\rm cont}$ significance at the location of the QSO in \jzotto\ and \jsedici. We also detect rest-frame $\sim125$ GHz continuum in the JVLA maps of \jzdue\ and \jzotto\ at $\sim8\sigma_{\rm cont}$ and $\sim9\sigma_{\rm cont}$ significance, respectively. 
Measured continuum flux densities range from $\sim~0.04$ mJy at rest-frame 125 GHz to $\sim~0.1$~mJy at 460 GHz and $\sim0.6$ mJy at 580 GHz (Table \ref{table:co-properties}). For the WISSH QSO \jquatt\ we measure a strong rest-frame 1935 GHz continuum flux density of $7.7\pm0.3$ mJy. The latter is a factor of $\sim10$ higher than the 1935 GHz continuum flux density found for the WISSH QSO J1015$+$0020 at $z\sim4.4$ in \cite{Bischetti18}, suggesting a large variety of far-infrared continuum properties in our sample, similarly to what observed in $z\sim5-6$ QSOs \citep[e.g.][]{Trakhtenbrot17, Nguyen20}.

The limited angular resolution of most observations presented in this work hampers a detailed study of the continuum emission morphology in these QSOs. Continuum morphology is consistent with a point source in all targets except \jquatt\ and \jqnove\, for which we measure a deconvolved angular size of (0.31$\pm$0.02)$\times$(0.27$\pm$0.02) arcsec$^2$ and (0.69$\pm$0.19)$\times$(0.50$\pm$0.20) arcsec$^2$, corresponding to $\sim2.0\times1.8$ kpc and $\sim5.7\times4.2$ kpc$^2$, respectively. 

In the field of view of \jqnove, a strong continuum emitter is detected at $\sim2.4$ arcsec from the nucleus (Fig. \ref{fig:contmaps}e), associated with a companion source (see Sect. \ref{sect:linemaps}) whose continuum emission is comparable to that of the QSO host-galaxy, with a size of (0.74$\pm$0.22)$\times$(0.43$\pm$0.23) arcsec$^2$, corresponding to $\sim6.1\times3.6$ kpc$^2$.
Weak, elongated continuum emission can be also observed
along the QSO-companion direction, possibly linked to a tidal feature.
Also in \jzdue, the elongated continuum structure which can be observed at $\sim5\sigma_{\rm cont}$ in the north-east direction, up to $\sim5$ arcsec from the nucleus, reveals the presence of a continuum-emitting nearby source (see Sect. \ref{sect:linemaps}) whose continuum flux is about 45\% of that of the QSO.

\subsection{CO and [CII] emission}\label{sect:linemaps}
Figure \ref{fig:linemaps} shows the continuum-subtracted, velocity-integrated maps of CO and [CII] emission for the eight WISSH QSOs analysed in this work. The detection rate in mid$-J$ CO and [CII] emission is 100\% with significance of $\sim4.5-30\ \sigma_{\rm cold}$, where $\sigma_{\rm cold}$ is the rms sensitivity (also given in Fig. \ref{fig:linemaps}) of the velocity-integrated emission line maps. CO(1$-$0) is also detected at $\sim5\sigma_{\rm cold}$ in one of the two WISSH QSOs targeted by JVLA, namely \jzdue, while no significant CO(1$-$0) emission is observed in \jzotto.
In Fig. \ref{fig:linemaps}, the location of the maximum FIR continuum emission (or the optical position of the QSO, in the case of continuum-undetected sources \jqotto\ and \jsedici) is also shown by yellow (red) stars. 
For most sources, a small offset between continuum and CO or [CII] emission can be observed, typically smaller than 0.5 arcsec. The largest offset, that is $\sim1$ arcsec, is observed between the SDSS optical position and the peak of CO(4$-$3) emission in \jqotto, although consistent within the uncertainty on the optical coordinates. 

In addition to cold gas emission from the host-galaxies of the WISSH QSOs, CO and [CII] maps revealed the presence of several line emitters detected within an angular separation of $\sim0.8$ arcsec up to $\sim16$ arcsec from the nuclei, as shown by coloured contours in Fig. \ref{fig:linemaps}. Specifically, six out of eight WISSH QSOs analysed here show line emitters within the ALMA or NOEMA field of view, detected at $\gtrsim5\sigma_{\rm cold}$ significance. Two of them, namely Comp$_{\rm J0209}$ and Comp$_{\rm J1549}$, are also detected in continuum emission (Sect. \ref{sect:contmaps}). For each of these sources, the proximity to the QSO in terms of sky frequency of the line emission (corresponding to a velocity shift in the range $\sim40-1800$ km~s$^{-1}$) and angular separation suggests that they are companion CO and [CII] emitters, located at approximately the same redshift of the QSO. 

\begin{figure*}[htb]
    \centering
    \includegraphics[width=0.75\textwidth]{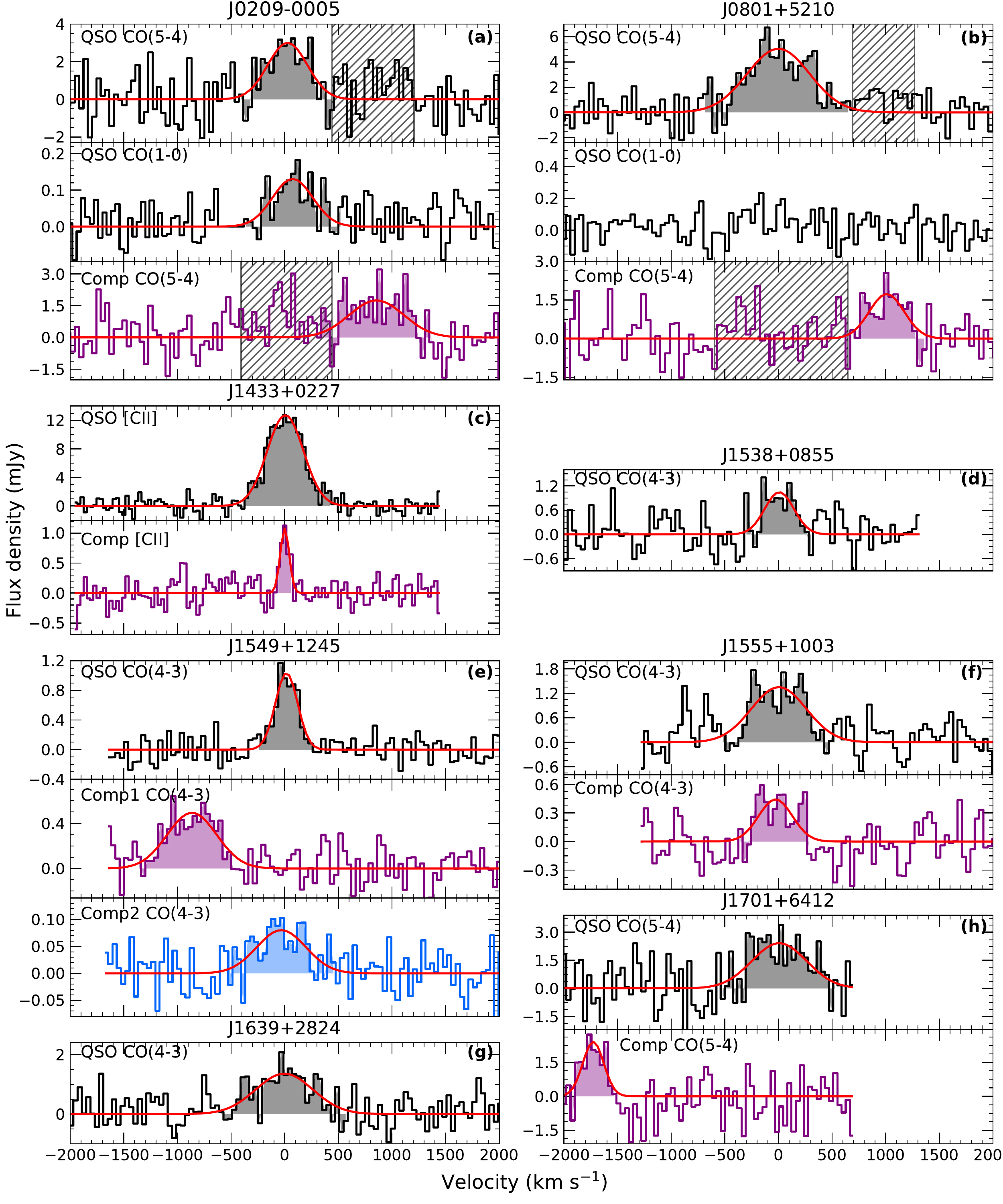}
    \caption{Spectra of the CO or [CII] emission lines for the WISSH QSOs analysed in this work. For each source, the targeted transition is indicated by the top label. Spectra were extracted from the region where emission from the QSO host-galaxy is detected with SNR$>2$ (see Fig. \ref{fig:linemaps}) and show the velocity range $v\in[-2000,+2000]$ \kms\ in channels of 40 \kms. In case of (e), we preferred a circular extraction region with size of the ALMA beam, centred on the QSO position, to limit contamination from a nearby line emitting source at 0.6 arcsec separation (see Fig. \ref{fig:linemaps}e). The best-fit, Gaussian profiles are shown by the red curve.  The filled histogram indicates velocity channels in which emission in the QSO and companion spectra is $>5\%$ than the peak flux of the best-fit model. In panels (a) and (b), grey spectral regions were excluded from the fit because associated with line emission from nearby sources (see Sect. \ref{sect:linemaps}) partially unresolved by the NOEMA beam.}
    \label{fig:spectra}
\end{figure*}

Fig. \ref{fig:spectra} shows the continuum-subtracted CO and [CII] spectra of the eight WISSH QSOs and of their companion galaxies, extracted from regions where line emission is detected at $>2\sigma_{\rm cold}$ in the QSO host-galaxy. For each source, the best-fit Gaussian model of the line is also displayed by the red curve. The redshift inferred from the Gaussian centroid ($z_{\rm cold}$), the FWHM and integrated flux density for each line are listed in Table \ref{table:co-properties}. Details about individual QSOs are given below. 

{\it \jzdue}: this QSO has been detected in both CO(5$-$4) and CO(1$-$0) emission in the NOEMA and JVLA maps (Fig. \ref{fig:linemaps}a). At the resolution and sensitivity of our observations, both transitions show a morphology which is consistent with a point source, although a low-significance tail can be observed in CO(1$-$0) south of the QSO location. Redshift and line width estimates based on CO(5$-$4) and CO(1$-$0) are consistent within the uncertainties. A strong CO(5$-$4) emitting companion is located north-east of the QSO at an angular separation of $\sim4$ arcsec, corresponding to a projected distance of $\sim32$ kpc at the redshift of the QSO. It shows a broad CO(5$-$4) profile, redshifted by $\sim800$ \kms\ with respect to the QSO. Continuum emission associated with the companion can be also seen at $\sim150$ GHz in the observed frame, as the elongated structure in Fig. \ref{fig:contmaps}a.

{\it \jzotto}: although characterised by a bright CO(5$-$4) emission with a peak flux density of 5 mJy, this source shows no detected CO(1$-$0) in the JVLA map (Fig. \ref{fig:linemaps}b). To infer an upper limit on the CO(1$-$0) integrated flux density S$\Delta v_{\rm 1-0}$, we have integrated the JVLA datacubes around $z_{\rm cold}$ over 685 \kms, assuming the same FWHM of CO(5$-$4). We have then computed S$\Delta v_{\rm 1-0}=3\times\sigma_{\rm CO, 1-0}$, where $\sigma_{\rm CO, 1-0}$ is the rms sensitivity of the velocity-integrated CO(1$-$0) map (Table \ref{table:co-properties}). 
In the NOEMA map, a nearby source emitting in CO(5$-$4) can be also seen at a projected distance of $\sim40$ kpc from \jzotto.

{\it \jquatt}: this QSO at $z_{\rm cold}=4.728$ shows a strong [CII] line with an integrated flux density of $5.40\pm0.24$ Jy \kms, which is consistent within the errors to the value of 4.79$\pm$0.38 Jy \kms\ found for this source by \cite{Nguyen20} for the same observation. [CII] emission in \jquatt\ is a factor of $\sim10$ brighter than what found in the WISSH QSO \jdieci\ at $z_{\rm cold}\sim4.4$, also targeted in [CII] by \cite{Bischetti18}. This result is consistent with the variety of [CII] properties typically observed in high-$z$ QSOs \citep[e.g.][]{Trakhtenbrot17,Nguyen20}. [CII] emission in \jquatt\ is spatially resolved, with a deconvolved size of (0.44$\pm$0.03)$\times$(0.38$\pm$0.03) arcsec$^2$, corresponding to $\sim2.9\times2.5$ kpc$^2$. A second weak (peak flux density $\sim1$ mJy) [CII] emitter has been detected at a distance of 12 kpc (Figure \ref{fig:linemaps}c), showing an almost null velocity shift of $60\pm40$ \kms\ with respect to the QSO redshift.

{\it \jqotto}: we have detected CO(4$-$3) emission with a modest significance of $\sim4.5\sigma_{\rm CO}$ from the host-galaxy of this QSO, located at $z_{\rm cold}\sim3.572$. The peak of CO emission is separated by $\sim1$ arcsec from the SDSS optical position of the QSO, the offset being comparable to the uncertainty on the SDSS position.  The line profile has a peak flux density of $\sim1$ mJy and a FWHM of 400$\pm40$ \kms. No additional sources have been detected in the ALMA maps. 

\begin{table*}[htb]
    \centering
    \caption{Properties of continuum and line (CO or [CII]) emission for the WISSH QSOs analysed here. Columns give the following information: (1) SDSS ID, (2) observed transition, (3) redshift estimated from (sub-)mm emission lines with typical uncertainty $\Delta z_{\rm cold}\sim0.001$, (4-5) FWHM and integrated flux of CO or [CII], (6) CO luminosity, (7) [CII] luminosity and (8) continuum flux. {Uncertainties do not include errors on abso}}
    \setlength{\tabcolsep}{4 pt}
    \begin{tabular}{lccccccc}
    \toprule
    ID  & Transition & $z_{\rm cold}$ & FWHM & $S\Delta$v & $L^\prime_{\rm CO}$ & $L_{\rm [CII]}$ & $S_{\rm cont}$  \\
        &      &           &  \kms & Jy \kms    & {\small 10$^{10}$ K \kms\ pc$^2$} & 10$^9$ \lsun & mJy \\
    (1) & (2) & (3) & (4) & (5) & (6) & (7) & (8) \\
    \midrule
    \jzdue  & \begin{tabular}{c} CO(5$-$4) \\ CO(1$-$0) \end{tabular} & \begin{tabular}{c} 2.870 \\ 2.870 \end{tabular} &  \begin{tabular}{c} 435$\pm$95\\ 440$\pm$85 \end{tabular} & \begin{tabular}{c} 1.4$\pm$0.3 \\ 0.062$\pm$0.012 \end{tabular}  & \begin{tabular}{c} 2.15$\pm$0.46 \\ 2.39$\pm$0.46 \end{tabular} &  $-$   & \begin{tabular}{c} 0.38$\pm$0.04 \\ 0.039$\pm$0.005 \end{tabular} \\
    \jzotto & \begin{tabular}{c} CO(5$-$4) \\ CO(1$-$0)\end{tabular} & \begin{tabular}{c} 3.256 \\ $-$ \end{tabular} & \begin{tabular}{c} 685$\pm$70 \\ $-$ \end{tabular} & \begin{tabular}{c} 3.68$\pm$0.40 \\ <0.095$^a$ \end{tabular} & \begin{tabular}{c} 6.99$\pm$0.76 \\ $<4.48$
    \end{tabular} &  $-$ & \begin{tabular}{c} <0.30$^{b}$ \\ 0.091$\pm$0.010 \end{tabular} \\
    \jquatt   & [CII]     & 4.728 &  400$\pm$40  & 5.40$\pm$0.24 &  $-$ & 3.72$\pm$0.16 & 7.7$\pm$0.3 \\
    \jqotto   & CO(4$-$3) & 3.572 &  320$\pm$90  & 0.36$\pm$0.11 & 1.23$\pm$0.38 & $-$ & <0.087$^{b}$ \\
    \jqnove   & CO(4$-$3) & 2.374 &  245$\pm$40  & 0.27$\pm$0.03 & 0.47$\pm$0.05 & $-$ & 0.12$\pm$0.01  \\
    \jqcinque & CO(4$-$3) & 3.529 &  605$\pm$90  & 0.87$\pm$0.14 & 2.94$\pm$0.68 & $-$ & 0.10$\pm$0.02 \\
    \jsedici  & CO(4$-$3) & 3.846 &  615$\pm$90  & 0.92$\pm$0.15 & 3.55$\pm$0.66 & $-$& <0.090$^{b}$ \\
    \jdsette  & CO(5$-$4) & 2.753 &  595$\pm$120 & 1.50$\pm$0.34 & 2.15$\pm$0.49 & $-$ & 0.60$\pm$0.06 \\
    J1015$+$0020$^*$ & [CII] & 4.407 &  340$\pm$40  & 0.47$\pm$0.05 &  $-$ & 0.29$\pm$0.03 & 0.60$\pm$0.06\\
    \midrule
    Comp$_{\rm J0209}$ & \begin{tabular}{c} CO(5$-$4) \\ CO(1$-$0) \end{tabular} & \begin{tabular}{c} 2.881 \\ $-$  \end{tabular} & \begin{tabular}{c} 600$\pm$95 \\ $-$  \end{tabular} & \begin{tabular}{c} 1.1$\pm$0.3 \\ $<0.044^{a}$ \end{tabular} & \begin{tabular}{c} 1.74$\pm$0.46 \\ $<1.71$ \end{tabular} & $-$ & \begin{tabular}{c} 0.14$\pm$0.03 \\ $<0.014^{b}$ \end{tabular} \\
    Comp$_{\rm J0801}$ & \begin{tabular}{c} CO(5$-$4) \\ CO(1$-$0) \end{tabular} & \begin{tabular}{c} 3.271 \\ $-$ \end{tabular} & \begin{tabular}{c} 385$\pm$65 \\ $-$ \end{tabular}& \begin{tabular}{c} 0.71$\pm$0.26 \\ $<0.058^{a}$ \end{tabular} & \begin{tabular}{c} 1.36$\pm$0.49 \\ $<2.78$ \end{tabular} & $-$ & \begin{tabular}{c} $<0.30^{b}$ \\ $<0.30^{b}$ \end{tabular} \\
    Comp$_{\rm J1433}$ & [CII] & 4.728 & 100$\pm$43 & 0.11$\pm$0.02 & $-$ & 0.08$\pm$0.02 & <0.015$^{b}$\\
    Comp1$_{\rm J1549}$ & CO(4$-$3) & 2.363 & 540$\pm$95& 0.29$\pm$0.05 & 0.49$\pm$0.08 & $-$ & 0.12$\pm0.01$ \\
    Comp2$_{\rm J1549}$ & CO(4$-3$) & 2.374 & 540$\pm$110 & 0.046$\pm$0.013 & 0.08$\pm$0.02 & $-$ & <0.021$^{b}$\\
    Comp$_{\rm J1555}$ & CO(4$-3$) & 3.531 & 370$\pm$70 & 0.29$\pm$0.06 & 0.98$\pm$0.20 & $-$ & <0.084$^{b}$  \\
    Comp$_{\rm J1701}$ &  CO(5$-$4) & 2.753 & 130$\pm$60 & 0.50$\pm$0.19 & 0.84$\pm$0.27 & $-$ & <0.30$^{b}$\\ 
    \bottomrule
    \end{tabular}
    \label{table:co-properties}
    \flushleft 
    \footnotesize $^*$ presented in \cite{Bischetti18}\\
    $^a$ upper limit computed for a point source as $3\times\sigma_{\rm cold}$, where $\sigma_{\rm cold}$ is the rms of the velocity integrated JVLA map produced by assuming the same FWHM of CO(5$-$4)\\
    $^{b}$ upper limit computed for a point source as $3\times\sigma_{\rm cont}$ (Table \ref{table:sample}).
\end{table*}

{\it \jqnove}: CO(4$-$3) emission in this QSO is characterised by a narrow line profile, with a FWHM=245$\pm$50 \kms, and a peak flux density of $\sim1$ mJy. Despite the modest resolution of ALMA observations (1 arcsec), CO emission is resolved and has a deconvolved size of (1.10$\pm$0.16)$\times$(0.86$\pm$0.16) arcsec$^2$. Indeed, \jqnove\ is a peculiar case, showing two line emitting sources within a small separation of $\sim2.4$ arcsec (Figure \ref{fig:linemaps}e). The strongest one, located at $\sim19$ kpc south-east of the nucleus, shows CO(4$-$3) emission as bright as that of the QSO, with a deconvolved size of (1.24$\pm$0.15)$\times$(0.55$\pm$0.13) arcsec$^2$. Such emission is blue-shifted by $\sim950$ \kms\ with respect to the redshift of \jqnove\ and shows a FWHM$\sim540$ \kms. A second, fainter CO(4$-$3) emitting companion is also observed $\sim12$ kpc north of the QSO, with a modest velocity shift of $200\pm40$ \kms.

{\it \jqcinque}: this QSO shows a broad (FWHM=615$\pm$90 \kms) CO(4$-$3) emission line profile peaking at $\sim1.4$ mJy, located at $z_{\rm cold}\sim3.529$. The resolution of our ALMA observations has allowed us to marginally resolve CO(4$-$3) emission produced by a nearby companion galaxy, distant only 6 kpc from the QSO in the west direction (Fig. \ref{fig:linemaps}f).

{\it \jsedici}: for this QSO, we measure a CO(4$-$3) based redshift $z_{\rm cold} = 3.846\pm0.001$, significantly larger than the redshift $z=3.840$ derived from the same data by \cite{Schramm19}. However, we note that the latter would correspond to an observed $\nu_{\rm obs}\sim95.37$ GHz, inconsistent with the CO(4$-3$) spectrum shown in their Fig. 3. The CO(4$-3$) emission line profile has a FWHM=$615\pm90$ \kms, and an integrated flux density of $\sim0.9$ Jy \kms. By fitting a 2D elliptical Gaussian to the data (in the image and uv plane), we measure an emission size  of (0.15$\pm$0.02)$\times$(0.11$\pm$0.02) arcsec$^2$ (Fig. \ref{fig:linemaps}g). We note that our measured minor axis is a factor of about two larger than the value reported by \cite{Schramm19}, likely due to the larger velocity range that we used to produce the velocity-integrated CO(4$-$3) map (solid histogram in Fig. \ref{fig:spectra}g), compared to \cite{Schramm19}, who reported a FWHM=$495\pm30$ \kms. No additional sources have been detected in the ALMA maps. 

{\it \jdsette}: we have detected CO(5$-4$) emission from this QSO, associated with a $z_{\rm cold}=2.753$. Similarly to the other CO(5$-4$) targets in our sample, the line peaks at $>2$mJy, with an associated integrated flux density of $1.5\pm0.35$ Jy \kms. In the NOEMA maps we have detected a companion galaxy distant $\sim130$ kpc from \jdsette, showing CO(5$-4$) blue-shifted (by $\sim1800$ \kms) emission, whose flux is about one third of that of the QSO.

For each target and for the companion galaxies, we compute CO or [CII] luminosity according to Eq. (3) and Eq. (1) in \cite{Solomon05}, respectively. The resulting CO luminosities, listed in Table \ref{table:co-properties}, are in the range $L^\prime_{\rm CO}\sim(0.5-7)\times10^{10}$ K km s$^{-1}$ pc$^2$ for the QSO host galaxies. For the companion galaxies, we measure $L^\prime_{\rm CO}$ in the range $(0.1-1.7)\times10^{10}$ K km s$^{-1}$ pc$^2$. The CO luminosity of the companion is typically a factor of $\sim3-5$ lower or, in the case of \jzdue\ and \jqnove, comparable to that of the QSO host.
In the case of \jquatt, we measure a [CII] luminosity  $L_{\rm [CII]}=(3.72\pm0.16)\times10^9$ \lsun.

\subsection{Investigating the CO spectral line energy distribution}\label{sect:sled}

\begin{figure}[thb]
    \centering
    \includegraphics[width =1\columnwidth]{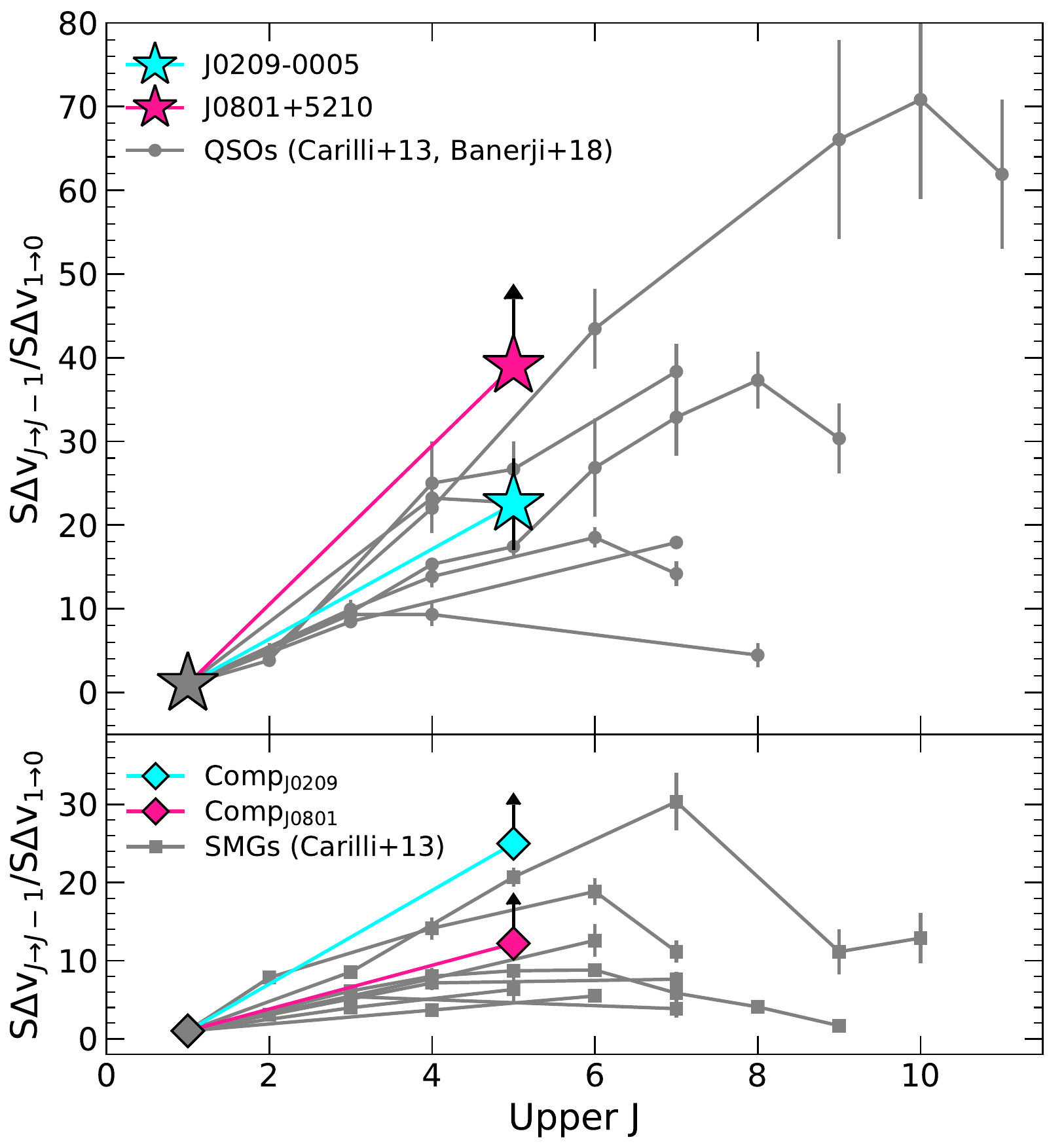}
    \caption{Top panel shows the CO SLED of  \jzdue\ and \jzotto\ with CO(1$-$0) and CO(5$-$4) observations, compared to QSOs from literature with available measure of the ground transition \citep{Carilli&Walter13}. Bottom panel displays the CO ladder of companion galaxies Comp$_{\rm J0209}$ and Comp$_{\rm J0801}$, together with CO SLEDs of SMGs. }
    \label{fig:cosled}
\end{figure}

As mentioned in Sect. \ref{sect:linemaps}, \jzdue\ and \jzotto\ have been observed in both CO(5$-$4) and CO(1$-$0) rotational transitions. 
It is therefore possible to derive some information on the excitation conditions of the gas in WISSH QSOs from the relative strength of these two lines. In the case of \jzdue, we measure a ratio of the integrated flux densities $S\Delta v_{5-4}/S \Delta v_{1-0}=22.5\pm6.0$, while for \jzotto\ we infer the lower limit $S\Delta v_{5-4}/S \Delta v_{1-0}>36.0$. 
Figure \ref{fig:cosled} shows the CO SLED of \jzdue\ and \jzotto, compared to few QSOs from literature with accurate measure of CO transitions up to $J\gtrsim5$ \citep{Carilli&Walter13}. We also include the $z\sim2.5$ reddened QSO ULAS J1234$+$0907 from \cite{Banerji18}.
We note that the $S\Delta v_{5-4}/S \Delta v_{1-0}$ ratio measured for \jzdue\ is similar to the values reported in other QSOs, while that of \jzotto\ suggests that CO can be highly excited in our hyper-luminous QSOs (see Sect. \ref{sect:disc-sled} for further discussion).

By translating CO flux ratios in CO luminosity ratios, we obtain $L^\prime_{\rm CO(5-4)}/L^\prime_{\rm CO(1-0)}=0.90\pm0.24$ and $L^\prime_{\rm CO(5-4)}/L^\prime_{\rm CO(1-0)}>1.39$ for \jzdue\ and \jzotto, respectively.
The ratio measured for \jzdue, based on the detection of both CO(1$-0$) and CO(5$-$4), is consistent within the uncertainty with the typical luminosity ratio reported for QSO host galaxies by \cite{Carilli&Walter13}. We thus adopt $L^\prime_{\rm CO(5-4)}/L^\prime_{\rm CO(1-0)}\simeq0.69$ and $L^\prime_{\rm CO(4-3)}/L^\prime_{\rm CO(1-0)}\simeq0.87$ \citep{Carilli&Walter13} to infer CO(1$-$0) luminosity in the remaining WISSH QSOs from the observed mid$-J$ CO rotational transitions. We note that by assuming the CO SLED of \jzdue, the inferred $L^\prime_{\rm CO(1-0)}$ would be lower by a factor of $\gtrsim2$. On the other hand, by considering the CO ladder of the Cloverleaf QSO \citep{Riechers11clover}, which shows the minimum measured $CO(5-4)/CO(1-0)$ ratio in Fig. \ref{fig:cosled}, the inferred $L^\prime_{\rm CO(1-0)}$ would be higher by a factor of $\sim1.3$.

Finally, the CO(5$-$4) and CO(1$-$0) observations also probe molecular gas excitation in the companion galaxies around \jzdue\ and \jzotto\ (see Sect. \ref{sect:linemaps}). The lower limits inferred for the integrated CO(5$-$4)/CO(1$-$0) flux ratios in Comp$_{\rm J0209}$ and Comp$_{\rm J0801}$ are shown in the bottom panel of Fig. \ref{fig:cosled}, compared with CO SLEDs of sub-millimetre galaxies (SMGs) from literature \citep{Carilli&Walter13}. The $S\Delta v_{5-4}/S \Delta v_{1-0}>25$ measured in Comp$_{\rm J0209}$ indicates a high CO excitation, similar to the excitation in the QSO host-galaxy, while in Comp$_{\rm J0801}$ we found $S\Delta v_{5-4}/S \Delta v_{1-0}>12$, consistent with the typical CO SLED of SMGs \citep[e.g.][]{Casey14}.

\section{Broad band SEDs and millimetre emission}\label{sect:sed}

The observations presented in Sect. \ref{sect:contmaps} have provided us with a measure of the rest-frame 430$-$1935 GHz continuum, which can be used to extend to (sub-)millimetre wavelengths the coverage of the spectral energy distribution (SED) presented in \cite{Duras17} for the half of our sample with available \textit{Herschel} photometry.  For QSOs with no \textit{Herschel} coverage, we combine ALMA and NOEMA continuum measurements with the average infrared properties of WISSH-\textit{Herschel} QSOs to provide an estimate of $L_{\rm IR}$.
\begin{figure}[]
    \centering
    \includegraphics[width = 1\columnwidth]{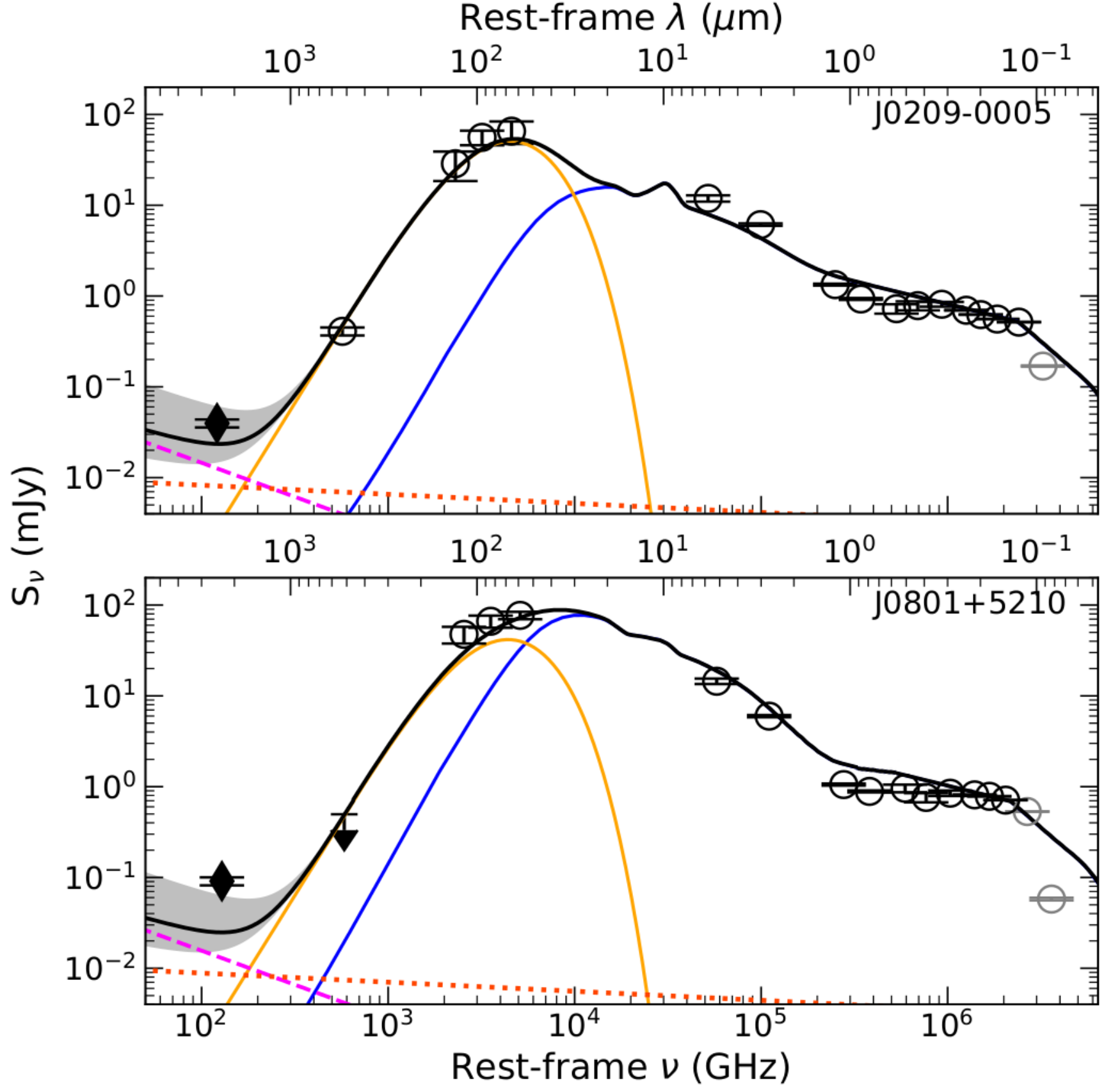}
    \caption{Rest-frame SED of \jzdue\ (top) and \jzotto\ (bottom). In each panel, black circles indicate the photometric points considered (arrows represent $3\sigma$ upper limits). Photometric points at $\lambda<1216 \AA$ are not included in the fits due to Ly$\alpha$ absorption (grey circles). Rest-frame 125 GHz continuum data (not included in the fit) are shown by diamonds. Black curve represents the total best fit model, while blue(orange) curve refers to the QSO(cold dust) emission component. Synchrotron and free free emission are shown by the dashed and dotted lines, respectively.}
    \label{fig:sed-radio}
\end{figure}
We complement sub-mm data with archival SDSS DR12 \citep{Alam15}, 2MASS \citep{Skrutskie06}, WISE \citep{Wright10}, and {\it Herschel}/SPIRE \citep{Pilbratt10,Griffin10} broad-band photometry. For target \jqcinque, undetected by 2MASS, we include photometric points obtained from dedicated observations at the ESO-INAF Rapid Eye Mount (\textit{REM}) telescope (P.I. M. Bischetti) and INAF Telescopio Nazionale Galileo (\textit{TNG}, P.I. V. Testa). These observations provided us with magnitudes $J= 17.4\pm0.11$, $H = 16.47\pm0.12$ and $K^\prime = 16.16\pm0.13$ for \jqcinque.

ALMA has revealed the presence of a strong continuum emitter close to \jqnove\ (Fig. \ref{fig:contmaps}). QSO and companion being separated by $\sim2.4$ arcsec, cannot be resolved as distinct objects in the {\it WISE} images, given the PSF, which is in the range $6.1-12$ arcsec. However, we have verified that emission is centred at the QSO location in all WISE bands and estimated the companion contamination, as traced by irregular morphology of the WISE contours, to be $<10 \%$. Moreover, the \jqnove\ companion is undetected in one hour exposure {\it VLT}/SINFONI H and K band observations of the {\it SUPER} ESO large program 196.A-0377  \citep[P.I. V. Mainieri,][Kakkad et al. 2020 A\&A accepted]{Circosta18}. An upper limit of $S_{\rm cont}<0.54$ mJy on the QSO continuum emission at $\sim785$ GHz inferred from ACA observations has also been included in the SED (Fig. \ref{fig:sed}). 

In the case of \jzdue,  the continuum emission associated with the companion galaxy is $\sim25-35\%$ of the QSO continuum in the NOEMA band (Fig. \ref{fig:contmaps}a). To compute the SED of \jzdue, we assume \textit{Herschel} photometry to be similarly dominated by emission from the QSO host-galaxy.

\begin{figure*}[htb]
\centering
\includegraphics[width = 1\linewidth]{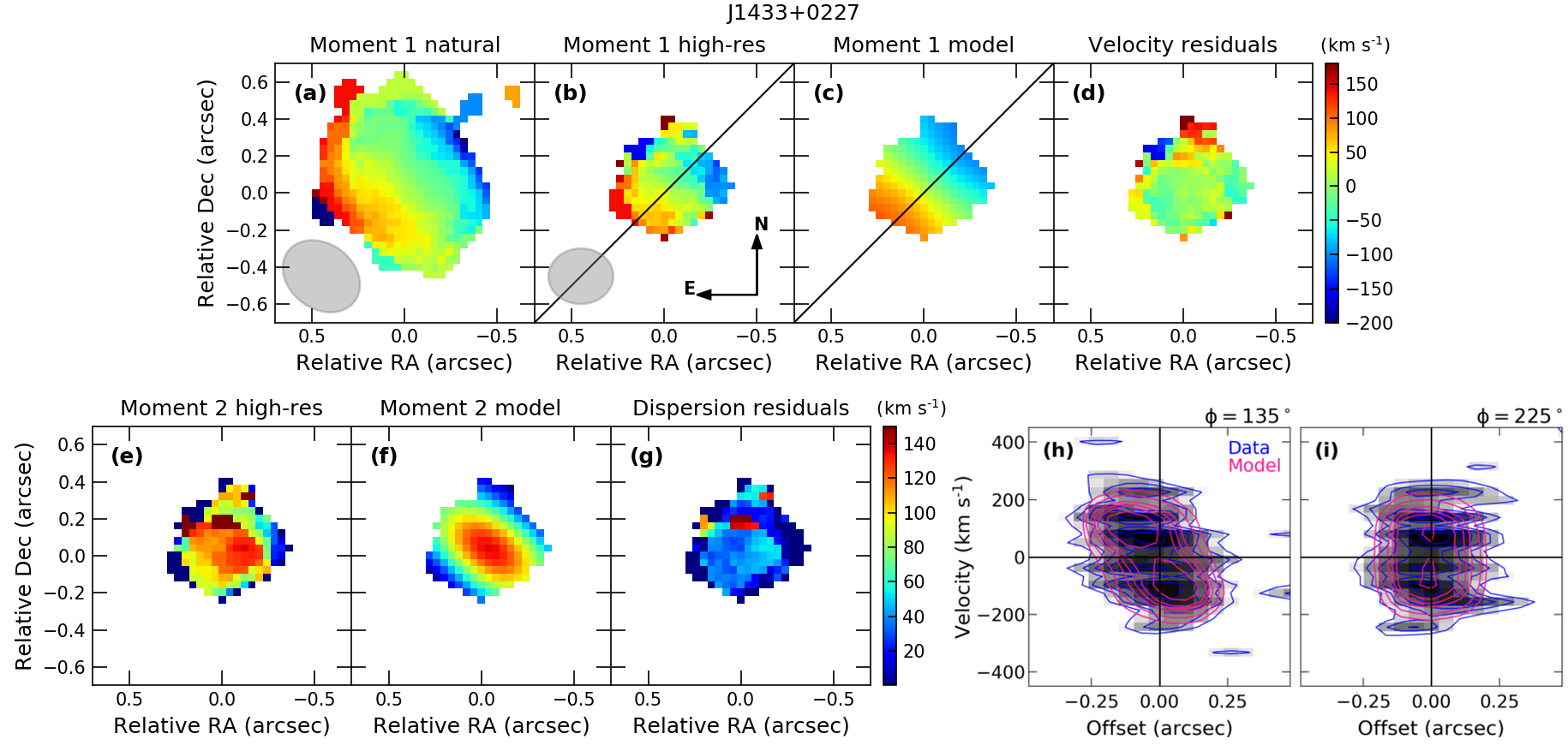}
\caption{Velocity maps of \jquatt\ associated with CO(4$-$3) emission detected at $>3\sigma_{\rm 40\ km s^{-1}}$ significance in the original (a) and high-resolution (b) maps, in which $\sigma_{\rm 40\ km s^{-1}}$= 0.43 mJy beam$^{-1}$ and 0.71 mJy beam$^{-1}$, respectively. ALMA beams are shown by the grey ellipses. Panel (c) displays the velocity map for the best-fit BAROLO model of a rotating disk. Black line indicates the major kinematic axis of the model. Velocity residuals are shown in panel (d). Panels (e)-(g) display the velocity dispersion maps associated with the high-resolution ALMA data and the best-fit BAROLO model, and the dispersion residual map. Position-velocity diagram associated with the high-resolution datacube along major (h) and minor (i) kinematic axes are also shown. Top labels indicate the associated position angles (measured anti-clockwise from north). Blue contours refer to the [2,3,4,5,6,8,...]$\sigma_{\rm 40\ km s^{-1}}$ significance.}
\label{fig:1433-moments}
\end{figure*}

SED Fitting was performed by using the procedure presented in \cite{Duras17,Zappacosta18} with a combination of QSO and host-galaxy emission components. In summary, the QSO component is described as the superposition of accretion disk emission \citep{Feltre12}, and of radiation coming from a clumpy, two-phase dusty torus \citep{Stalevski16}. Cold dust emission is modelled as a non-approximated modified blackbody that accounts for emission powered by star formation, absorbed and then re-emitted by the surrounding dust at mid- and far-infrared wavelengths. We use a library of 70 modified black-body templates covering the temperature range from 30 to 100 K, and we assume a power-law dependence of the optical depth with wavelength, that is $\tau_\lambda\propto\lambda^\beta$, where $\beta=1.6$ is the dust emissivity index typically used in high-$z$ QSOs \citep[e.g.][]{Beelen06}.
Dust temperature is a free parameter for all sources with \textit{Herschel} photometry, while for the remaining sources we assume $T_{\rm dust}=60$ K, i.e. the average measured temperature for the total sample of 16 WISSH-\textit{Herschel} QSOs \citep{Duras17}. We note that assuming a $T_{\rm dust}=40(80)$ K would result in a 50\% lower(a factor of two higher) $L_{\rm IR}$.

Fig. \ref{fig:sed-radio} shows the resulting rest-frame UV to far-infrared SEDs for \jzdue\ and \jzotto, while the SEDs of the other WISSH QSOs are shown in the Appendix. For all sources, the SED is dominated by the QSO component (blue curve) at all wavelengths below several tens of \mum. By integrating the QSO component in the range $1-1000$ \mum\ we infer intrinsic QSO bolometric luminosities in the range \lbol$\sim3.6\times10^{47}-1.0\times10^{48}$ \ergs. Concerning emission from cold-dust in the QSO host-galaxies (orange curve), we measure large values of the infrared, $8-1000$ \mum\ integrated luminosity in our targets, that is $L_{\rm IR}\simeq(1.7-8.3)\times10^{46}$ \ergs\ (Table \ref{table:inferred}).
However, some authors \citep[e.g.][]{Symeonidis17, Kirkpatrick19} argued that a significant fraction of the infrared emission in luminous AGN might be due to dust heated by the large radiative output of the accreting SMBH up to kpc scale. In particular, the amount of AGN-heated dust in the WISSH-\textit{Herschel} QSOs was estimated through radiative transfer by \cite{Duras17}, who found an average AGN contribution to $L_{\rm IR}$ of 50\% at \lbol$>10^{47}$ \ergs. Accordingly, we compute the AGN-corrected infrared luminosities of our WISSH QSOs by dividing by a factor of two the values listed in Table \ref{table:inferred}. Therefore, from these corrected values we derive the SFR using the relation by \cite{Kennicutt98} scaled to a Chabrier initial mass function.

In Fig. \ref{fig:sed-radio}, the rest-frame $\sim125$ GHz continuum of \jzdue\ and \jzotto\ is shown (solid diamonds). These JVLA photometric points were not included in the fitting because non-thermal and thermal bremsstrahlung emission significantly contribute to the SED at these frequencies, in addition to dust emission. Disentangling the relative contribution of these processes is not feasible without several photometric points at $\nu<200$ GHz. However, it is possible to estimate the AGN contribution at these frequencies by comparing the observed JVLA fluxes with typical emission from star forming galaxies. Specifically, we use the (AGN-corrected) SFR derived from SED fitting to compute the synchrotron and free free energy distribution (dashed and dotted lines in Fig. \ref{fig:sed-radio}) according to Eq. 13 and Eq. 12 by \cite{Yun&Carilli02}, respectively.
The gray shaded area indicates the scatter of non-thermal emission observed in local starburst galaxies for a given SFR \citep{Yun&Carilli02}. In the case of \jzdue, the $\sim125$ GHz continuum is consistent with no AGN contribution, while in \jzotto\ the QSO continuum shows a small excess  (by factor of $\sim1.5$) with respect to the values found in the starburst case. This may suggest a small but non negligible contribution of the AGN to mm-radio wavelengths. The extrapolation at radio wavelengths of the synchrotron and free free SED of both \jzdue\ and \jzotto\ is consistent with the non detections at 1.4 GHz in the VLA Faint Images of the Radio Sky at Twenty-Centimeters (FIRST) survey \citep{Helfand15}.

\begin{SCfigure*}[3][htb]
    \caption{Velocity (left) and velocity dispersion (right) maps of \jqnove\ and its bright companion, associated with CO(4$-$3) emission detected at $>3\sigma$ significance. In both panels, top(bottom) colorbar labels indicate values observed in the QSO(companion). The ALMA beam is also shown by the grey ellipse.}
    \includegraphics[width=0.65\textwidth]{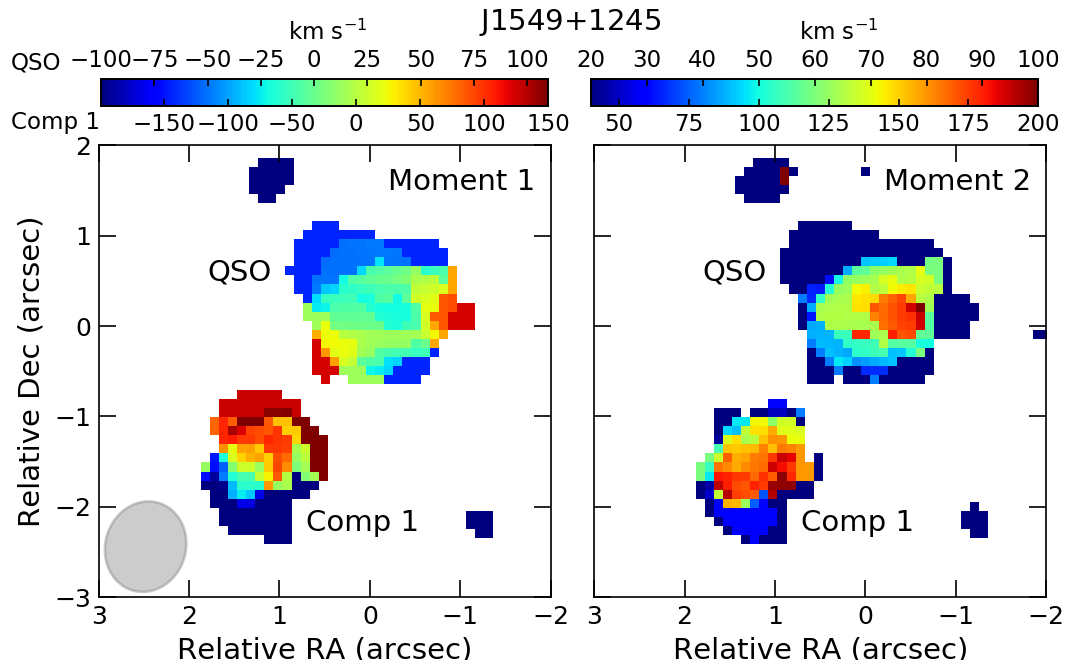}\hspace{0.2cm}
    \label{fig:1549-moments}
\end{SCfigure*}

\begin{SCfigure*}[3][htb]
    \caption{Velocity (left) and velocity dispersion (right) maps of \jsedici, associated with CO(4$-$3) emission detected at $>3\sigma$ significance. The ALMA beam is also shown by the grey ellipse.}
    \includegraphics[width=0.65\textwidth]{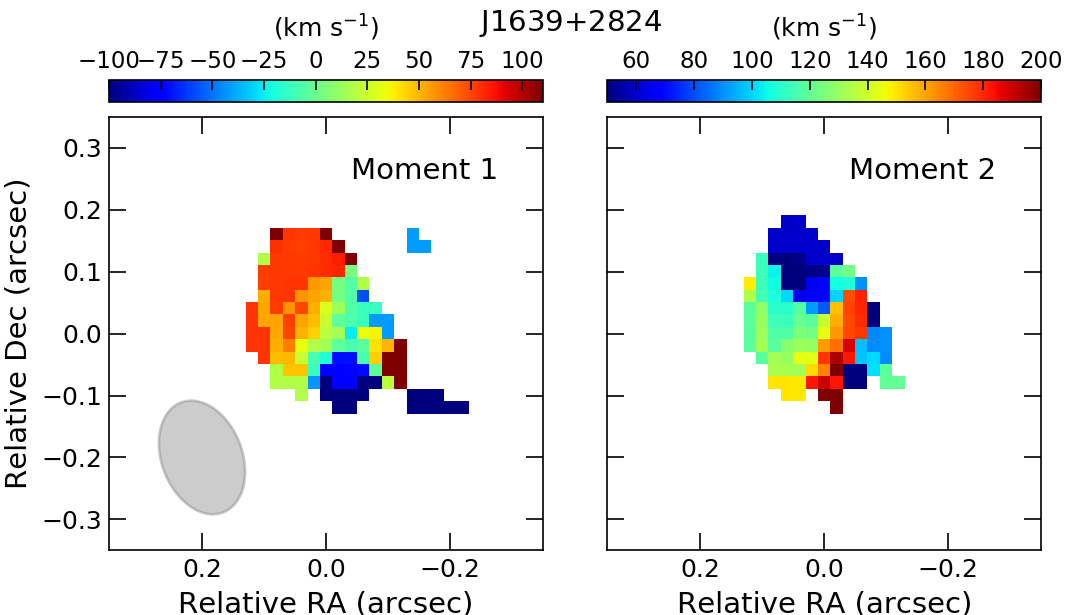}
    \label{fig:1639-moments}
\end{SCfigure*}

By exploiting the relation between the infrared continuum emission and the cold-dust mass, we derive an estimate of the dust content in the host-galaxy of the WISSH QSOs as follows \citep[e.g.][]{Beelen06, Scoville16}
\begin{equation}
    M_{\rm dust} = \frac{S_{\lambda}D_{L}^2}{k_{\rm dust}B(\lambda, T_{\rm dust})}
\end{equation}
where S$_\lambda$ is the continuum flux density at rest-frame $\lambda=850$ $\mu$m, measured from the best-fit SED model, $D_{\rm L}$ is the luminosity distance, $k_{\rm dust}=0.077$ kg$^{-1}$ m$^2$ is the dust mass opacity at 850 $\mu$m \citep{Dunne00} and $B(\lambda, T_{\rm dust})$ is the Planck function for a given dust temperature. The latter is also based on our SED fitting, in the range $T_{\rm dust}\in[40,75]$ K. The resulting $M_{\rm dust}$ are listed in Table \ref{table:inferred}.


\section{CO and [CII] kinematics}\label{sect:kine}

We analyse here the CO and [CII] kinematics of the three WISSH QSOs with spatially resolved emission, for which it has been possible to investigate the presence of gradients in the velocity (Moment 1) and velocity dispersion (Moment 2) maps.

Fig. \ref{fig:1433-moments}a shows the velocity map associated with [CII] emission in \jquatt. A velocity gradient is detected in approximately the north-west to south-east direction, with a projected velocity in the range $v\in[-100,+130]$ \kms, although the gradient associated with emission in the inner 0.5 arcsec around the nucleus shows a different orientation than emission at larger angular separation. Perturbations are present in the north direction, where emission with v$\sim0$ \kms\ smears out the gradient. 
We have fitted the 3D-BAROLO tilted-ring model \citep{DiTeodoro15} to the data to provide a description of the kinematics. However, the combination of disturbed morphology and limited angular resolution ($0.44\times0.34$ arcsec$^2$) hampers an accurate estimate of the source inclination and, in turn, of the kinematical properties. We have thus produced an ALMA [CII] datacube with increased angular resolution by applying a Briggs weighting of the visibilities with robust parameter $b=-0.5$ \citep{Briggs95}, resulting in a beamsize of $0.35\times0.30$ arcsec$^2$ (Fig. \ref{fig:1433-moments}b). The associated position-velocity diagram of the [CII] emission detected above 3$\sigma$ ($\sigma_{\rm 40 km s^{-1}}=0.71$ mJy beam$^{-1}$) is also shown in Fig. \ref{fig:1433-moments} h and Fig. \ref{fig:1433-moments}i.

\begin{table*}[htb]
    \centering
    \caption{Columns give the following information: (1) SDSS ID, (2) SED-based infrared luminosity (see Sect. \ref{sect:sed}) in the range $\lambda=8-1000$~$\mu$m. Typical uncertainty on $L_{\rm IR}$ is 0.1 dex for sources with {\it Herschel} photometry, while for the other QSOs we consider a 0.3 dex error. (3-4) AGN-corrected SFR and star-formation efficiency, (5) dust mass, (6) molecular gas mass, (7) gas-to-dust ratio, and (8) molecular gas depletion time. In the last row we report the same quantities measured for \jdieci, a WISSH QSO presented in \cite{Bischetti18}}
    \begin{tabular}{lccccccc}
    \toprule
    ID  & log $L_{\rm IR}$  &  SFR    &  SFE  & log $M_{\rm dust}$ & log$M_{\rm gas}$ & $M_{\rm gas}/M_{\rm dust}$ & $\tau_{\rm dep}$\\
        &     \lsun         & \msunyr &  {\small \lsun/(K km s$^{-1}$} pc$^2$) & \msun  & \msun & & Myr  \\ 
    (1) & (2) & (3) & (4) & (5) & (6) & (7) & (8)\\
    \midrule
    \jzdue    & 13.17 & 735 [465$-$1165] & 310 [230$-$415] & 8.14 & 10.28$\pm$0.08 & 140 & 25 [15$-$40] \\
    \jzotto   & 13.24 & 875 [550$-$1385] & 195 [150$-$250] & 8.22 & 10.55$\pm$0.04 & 210 & 40 [25$-$65] \\
    \jquatt   & 12.74 & 270 [170$-$430] & $-$ & 8.74 & \hspace{0.15cm} 11.00$\pm$0.30$^{b}$& \ 180$^{a}$  & 360 [180$-$700] \\
    \jqotto   & 12.82 & 330 [165$-$660] & 235 [120$-$470] & 8.04 & 10.05$\pm0.12$ & 100 & 35 [15$-$65] \\
    \jqnove   & 12.52 & 165 [85$-$330] & 305 [155$-$610] & 7.64 & 9.64$\pm$0.04  & 100 & 25 [15$-$50] \\
    \jqcinque & 12.94 & 435 [220$-$870] & 130 [65$-$260]  & 8.01 & 10.43$\pm$0.09 & 260 & 60 [30$-$120] \\
    \jsedici  & 12.72 & 260 [130$-$520] & 65 [35$-$130]  & 8.04 & 10.51$\pm0.07$ & 300 & 125 [55$-$245] \\
    \jdsette  & 13.34 & 1095 [690$-$1735] & 460 [340$-$625] & 8.14 & 10.28$\pm0.09$ & 140 & 15 [10$-$25] \\
    J1015$+$0020$^*$ & 12.11 & 100 [65$-$160] & $-$ & 7.02  & \hspace{0.25cm}9.28$\pm$0.30$^{b}$ & \ 180$^{a}$ & 20 [10$-$45] \\
    \bottomrule
    \end{tabular}
    \label{table:inferred}
    \flushleft 
    \footnotesize $^*$ presented in \cite{Bischetti18}\\
    $^a$ average value measured from WISSH QSOs with CO-based $M_{\rm gas}$\\
    $^{b}$ computed assuming $M_{\rm gas}/M_{\rm dust} = 180$
\end{table*}

The 3D-BAROLO fit of the high-resolution datacube results into an inclination $i = 34\pm10$ deg, which is consistent with that derived from the projected axes ratio of the CO emission measured in Sect. \ref{sect:linemaps} and the $i\sim39$ deg reported by \cite{Nguyen20}.
The velocity residuals map (Fig. \ref{fig:1433-moments}d) globally shows small deviations comparable to the spectral resolution,  except for the northern region, where blue- and red-shifted residuals up to $|v|\sim150$ \kms can be observed. Residuals in correspondence of the northern region are also visible in the dispersion residuals map (Fig. \ref{fig:1433-moments}g). Similarly, the overlay of data and BAROLO contours in the position-velocity diagram indicates that the bulk of the [CII] emission in \jquatt\ is compatible with a rotating disk, although deviations from pure rotation are visible along both major and minor kinematic axes.
We infer an intrinsic circular velocity $v_{\rm circ}=1.3\times\Delta v_{\rm blue-red}/(2{\rm sin}i)\simeq400$ \kms\ \citep[e.g.][]{Forster-Schreiber09,Tacconi13,Bischetti19}, where $\Delta v_{\rm blue-red}\simeq350$ \kms\ is the total velocity gradient measured from contours of BAROLO model at $\sim5\sigma$ in the position-velocity diagram at a projected distance of $\sim1.2$ kpc from the nucleus.

Fig. \ref{fig:1549-moments} reports the velocity and velocity dispersion maps associated with CO(4$-3$) emission in the WISSH QSO \jqnove. At the resolution of $\sim1$ arcsec$^2$, we do not identify a clear velocity gradient, the bulk of the emission showing a flat velocity profile with $|v|<50$ \kms. Structures with blue-shifted  ($v\sim-150$ \kms) and red-shifted ($v\sim+100$ \kms) velocity are observed only in the outer region. The \jqnove\ velocity profile is consistent with the low inclination $i\sim40$ deg derived from the projected axes ratio.
The velocity dispersion map shows a peak dispersion value of $\sim100$ \kms\ about 0.5 arcsec offset from the QSO position in the west direction, which may be related to disturbed kinematics. Given the SNR$\sim16$ of the CO emission, increasing the angular resolution by Briggs weighting the visibilities does not provide further information on the \jqnove\ kinematics.
Fig. \ref{fig:1549-moments} also shows the moment maps of the bright companion south-east of \jqnove, namely Comp1$_{\rm J1549}$.
This source shows a large velocity gradient with $v\in[-200, +150]$ \kms, approximately in the south-east to north-west direction. Such gradient may be interpreted as the bulk of CO(4$-$3) in Comp1 originating from an inclined ($i\sim64$ deg) rotating disk, although the limited SNR and resolution do not allow us to fit a disk model to the data. The velocity dispersion map displays large values up to $\sigma\sim200$ \kms\ at the position of Comp1 and in the direction of \jqnove. 

Although the modest SNR of the CO(4$-3$) emission prevents us from performing a BAROLO fit of the data, the velocity map of \jsedici\ (Fig. \ref{fig:1639-moments}) appears similar to that of \jquatt. Indeed, a velocity gradient is detected in the north-east to south-west direction, with projected velocity range $v\in[-100, +100]$ \kms, smeared out at $\sim[-0.1,-0.05]$ arcsec by some disturbances. The presence of perturbed motions is also supported by the high velocity dispersion of the gas in the same region, where $\sigma\sim150-200$ \kms. 

We point out that, in the case of marginally resolved targets observed with modest significance, such as \jqnove\ and \jsedici, the interpretation of moment maps is not straightforward and can be affected by large uncertainties. As an example, non-rotating ISM components such as outflows and merger-induced perturbations may be mixed with rotation and affect the observed velocity gradients \citep[e.g.][]{Diaz-Santos16,Trakhtenbrot17,Bischetti19}. Moreover, velocity dispersion profiles in the central regions may be affected by beam-smearing effects \citep[e.g.][]{Davies11,Tacconi13}. Higher angular resolution of these QSOs are needed to properly map the cold gas kinematics in their host galaxies.


\section{Discussion}\label{sect:discussion}

\subsection{The rich environment of hyper-luminous QSOs}

In a pilot study targeting [CII] emission around the WISSH QSO \jdieci, \cite{Bischetti18} found a large galaxy overdensity with three [CII] emitters at close distance ($\sim2-16$ kpc) and two continuum emitters likely physically associated.
In this work we build on the \cite{Bischetti18} result by testing the role of luminous QSOs as signposts of large galaxy overdensities in a fair statistical sample.

We confirm that QSOs at the brightest end of the luminosity function with $L_{\rm Bol}>10^{47}$ \ergs\ pinpoint high density regions, as we find that $\sim75\%$ ($80\%$, once \jdieci\ is included) of WISSH sources show at least one companion sub-millimetre galaxy. Specifically, six out of eight QSOs analysed here have a nearby line emitter, showing bright CO or [CII] emission at about the same redshift and in most cases of comparable luminosity to that of the QSO host-galaxy. These emitters are generally located at close distance ($\sim6-40$ kpc) from the QSO, with the exception of Comp$_{\rm J0209}$ which is located further away ($\sim130$ kpc). Two of them, namely Comp$_{\rm J0209}$ and Comp1$_{\rm J1549}$, are detected also in continuum emission, with a flux which is $\sim30-100\%$ that of the QSO. This suggests that a significant fraction of the SFR in the observed QSO-companion systems takes place in the companion galaxies, similarly to \cite{Bischetti18} findings and in agreement with expectations from semi-analytical models \citep[e.g.][]{Fogasy17}.

Our results provide useful observational evidence to the hotly debated issue of whether high-redshift QSOs reside in overdense regions.
Traditional studies of QSO environment mainly rely on {\it HST} imaging in the UV and optical bands. In the case of dust-reddened or Type 2 QSOs, a large fraction ($\sim50-80\%$) have been found to reside in interacting systems \citep{Urrutia08,Glikman15,Wylezalek16,Fan16}. However, in Type 1 sources such as WISSH QSOs, the overshining nuclear emission complicates the detection
of tidal features and close/minor mergers, leading to conflicting results about the fraction of interacting systems \citep[e.g.][]{Mechtley16}. 
In the last years, ALMA and NOEMA interferometers have expanded our knowledge of the environment around all types of QSOs by accessing a spectral regime that is mostly uncontaminated by AGN emission. Several works based on individual and small samples of high$-z$ QSOs reported the detection of nearby sources at sub-millimetre wavelengths, e.g. \cite{Ivison08,Clements09,Salome12,Riechers13,Fogasy17,Fogasy20,Diaz-Santos18}. The discovery of companion galaxies at sub-millimetre wavelengths has been pushed to the highest redshifts ($z\sim5-6$), revealing that $\sim15-50\%$ of luminous QSOs are accompanied by interacting companions  \citep[e.g.][]{Decarli17, Trakhtenbrot17,Dodorico18}. These companions can be as bright as the QSO host galaxies and significantly contribute to the mass growth of the host \citep[e.g.][see also Sect. \ref{sect:mdyn}]{Neeleman19}.
The observed high fraction of WISSH QSOs with nearby companions supports previous studies of the merger fraction as a function of AGN luminosity  \cite[e.g.][]{Treister12}, measuring a value close to unity in bright QSOs in agreement with a major role of mergers in triggering nuclear activity. Indeed, the merger fraction appears to steeply increase with redshift (by a factor of three from $z\sim1$ to $z\sim2.5$) in massive galaxies such as WISSH hosts \citep[e.g.][]{Conselice14}.
We note that the modest angular resolution and sensitivity of our observations (see Table \ref{table:sample}) prevent us from assessing the presence of an ongoing merger phase in all galaxies with a detected companion galaxy. Indeed, this will require to accurately probe disturbed ISM kinematics and to detect tidal features with higher angular resolution follow-up observations \citep[e.g. see][]{Diaz-Santos18,Decarli19}. 


\subsection{Star formation efficiency and gas depletion}\label{sect:SFE}

\begin{figure*}[htb]
    \centering
    \includegraphics[width = 1\linewidth]{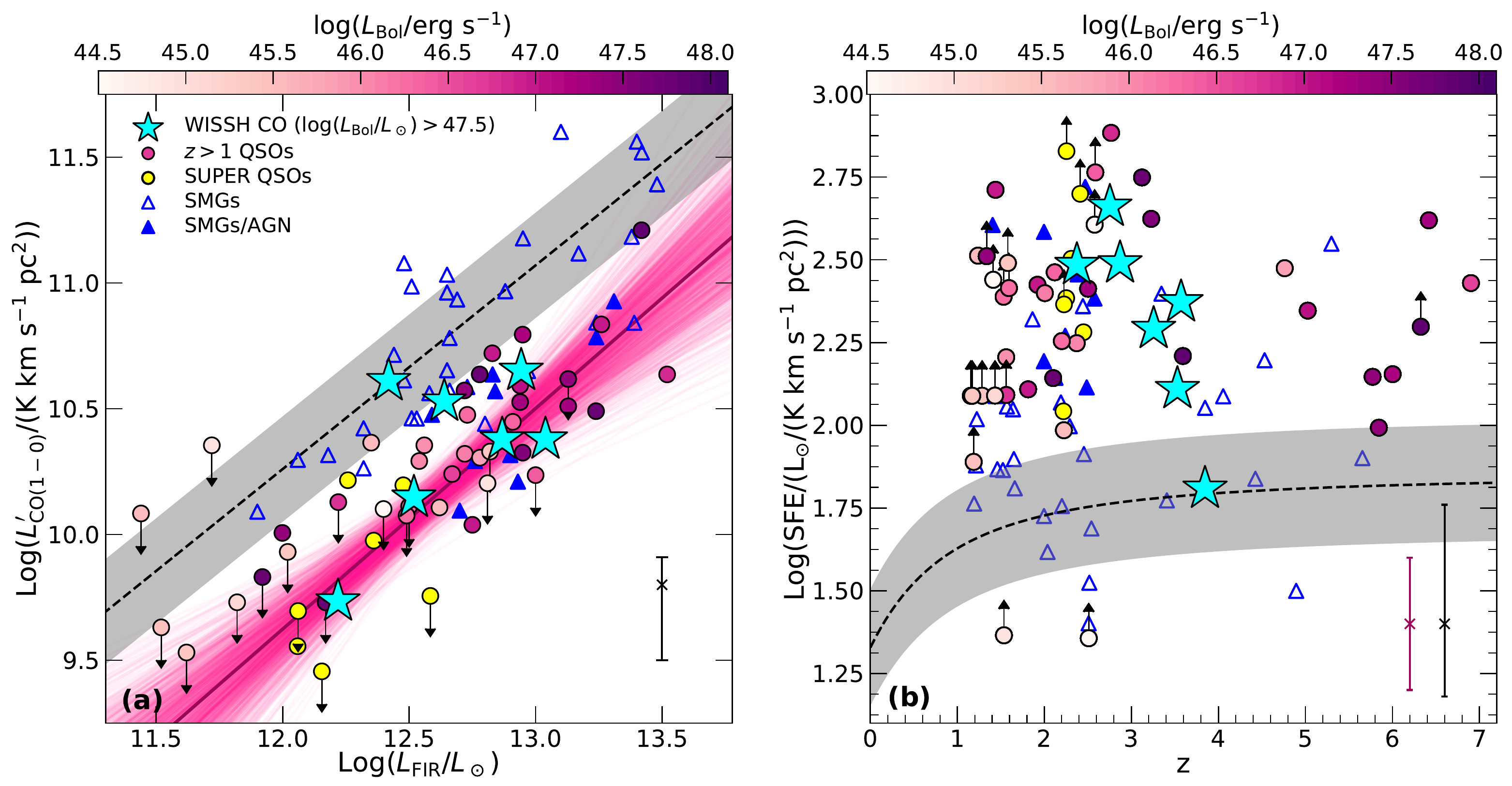}
    \caption{Panel (a) displays the CO(1$-0$) luminosity as a function of the infrared luminosity for the WISSH QSOs analysed in this work (cyan stars), compared with a sample of QSOs and SMGs from literature (see text for details). We show the best-fit $L^\prime_{\rm CO(1-0)}$ vs $L_{\rm IR}$ relation measured for the QSOs sample (solid line), as compared to the relation found for MS galaxies by \cite{Sargent14,Speagle14} (dashed line), characterised by a 0.2 dex scatter (grey shaded area). A thousand random realisations from our Bayesian fit are also shown (pink lines). Top colorbar indicates the AGN bolometric luminosity for the $z>1$ QSOs collection from literature. Error bar indicates the uncertainty on $L^{\prime}_{\rm CO(1-0)}$ associated with the assumption of a CO SLED, given the range of  $CO(5-4)/CO(1-0)$ ratios measured in high-z QSOs (Fig. \ref{fig:cosled}). Panel (b) shows the star formation efficiency as a function of redshift, compared to the SFE of MS galaxies with $M_{\*}\sim10^{11}$ M$_{\odot}$ (black line with grey area). Black(purple) error bar indicates the typical uncertainty on SFE for WISSH targets including(without including) the systematic error on the assumption of a CO SLED (see Sect. \ref{sect:sled}).}
    \label{fig:lcolfir.pdf}
\end{figure*}

We investigate here the relation between CO(1$-$0) luminosity, tracer of the molecular gas content \citep[e.g.][]{Bolatto13}, and the IR luminosity, tracer of the dust-reprocessed star formation rate \citep[e.g.][]{Kennicutt12} in the host-galaxy of the WISSH QSOs. Figure \ref{fig:lcolfir.pdf} (left panel) displays the location of our targets in the $L_{\rm IR}-L^{\prime}_{\rm CO(1-0)}$ plane, where $L_{\rm IR}$ is the AGN-corrected infrared luminosity (see Sect. \ref{sect:sed}). We show the best-fit relation derived for the compilation of $0<z\lesssim3$ main sequence galaxies by \cite{Sargent14}, indicated by the dashed line. As comparison samples, we include:
\begin{itemize}
\item[--] non-lensed, luminous SMGs at $z\sim1-4$ from \cite{Bothwell13}, starburst galaxies at $z\sim1.5-2.5$ from \cite{Silverman15,Yan10}, and dusty, highly star-forming galaxies at $z>4$ by \cite{Fudamoto17}, for a total of 40 galaxies. In the case of \cite{Bothwell13} targets, we distinguish between SMGs with and without AGN. 
\item[--] a subsample of 36, non-lensed $z>1$ QSOs with a good estimate of the infrared luminosity from the compilation by \cite{Perna18}. In particular, we consider QSOs with at least two photometric points in the rest-frame wavelength range 50-850 $\mu$m, and at least one detection of the far-infrared continuum within $50-200$ $\mu$m, that is close to the peak of the cold dust emission. This choice limits the systematics on $L_{\rm IR}$ due to different modelling techniques and assumptions on the infrared SED \citep[e.g.][]{Perna18,Kirkpatrick19}. Following the same approach used for our targets, we homogeneously compute $L^\prime_{\rm CO(1-0)}$ of all $z>1$ QSOs from the mid-$J$ CO luminosity reported by \cite{Perna18} according to the $L^{\prime}_{\rm CO(J\rightarrow J-1)}/L^{\prime}_{\rm CO(1-0)}$ ratios by \cite{Carilli&Walter13}, and derive the AGN-corrected $L_{\rm IR}$ following \cite{Duras17} for the $L_{\rm Bol}>10^{47}$ erg s$^{-1}$ sources. 
\item[--] X-ray selected QSOs at $z\sim2$ drawn from SUPER (SINFONI Survey for Unveiling the Physics and Effect of Radiative feedback), which have a reliable measure of $L_{\rm IR}$ \citep[][2020 in prep.]{Circosta18}. $L^\prime_{\rm CO(1-0)}$ and the AGN-corrected $L_{\rm IR}$ have been derived as in the previous point.
\end{itemize}

WISSH QSOs, with the exception of \jquatt, lie below the locus of high-$z$, main sequence galaxies (Fig.\ref{fig:lcolfir.pdf}). In general, the bulk of the $z>1$ QSO population shows lower CO luminosity than MS galaxies for a given $L_{\rm IR}$, with no clear dependence on the AGN bolometric luminosity. To quantify this difference, we fit data points of WISSH QSOs and the comparison QSOs sample with a linear regression based on a Bayesian approach, by using the python package \textit{linmix} \citep{Kelly07}. This allows us to consider errors on both variables $L_{\rm IR}$ and  $L^{\prime}_{\rm CO}$ and upper limits on the latter. For WISSH QSOs, we use the statistical uncertainty on $L_{\rm IR}$ from SED fitting ($\sim0.1$ dex), for sources with {\it Herschel} plus sub-mm photometric coverage, while we consider a $0.3$ dex systematic uncertainty for sources with no {\it Herschel} (Sect. \ref{sect:sed}). Errors on $L^{\prime}_{\rm CO}$ are listed in Table \ref{table:co-properties}. For the compilation of QSOs from literature, the typical uncertainties are 0.2 dex and 0.1 dex  on $L_{\rm IR}$ and $L^{\prime}_{\rm CO}$, respectively. The resulting best-fit correlation indicates that QSOs with infrared luminosity in the range $10^{12}-10^{13}$ \lsun\ (probed by WISSH QSOs and the majority of QSOs from literature) show on average a factor of $\sim4$ lower CO luminosity  than MS galaxies.
Differently, we note that the majority of SMGs follow the MS galaxies sequence, with sources  below the relation being mostly associated with the presence of an AGN.

Our results provide new insights into the location of QSO host-galaxies in the  $L^{\prime}_{\rm CO}-L_{\rm IR}$ plane \citep[e.g.][]{Genzel10,Carilli&Walter13}, and are similar to findings by \cite{Perna18}, who reported a reduced CO luminosity in high-$z$ QSOs for both obscured and unobscured sources. A different scenario was reported by \cite{Kirkpatrick19}, who performed a new, homogeneous analysis of most QSOs included in the \cite{Perna18} collection, finding no significant difference in the $L^{\prime}_{\rm CO}-L_{\rm IR}$ properties of QSOs and MS galaxies.  \cite{Kirkpatrick19} identified this discrepancy as mainly due to different SED fitting prescriptions at IR wavelengths and the use of a diverse CO SLED. By comparing estimates of $L_{\rm IR}$ and $L^{\prime}_{\rm CO(1-0)}$ for common sources between these two works, we find most $L_{\rm IR}$ values being consistent within a factor of two, while $L^{\prime}_{\rm CO(1-0)}$ is systematically higher in \cite{Kirkpatrick19} by a factor of $\sim1.5-2$, and up to a factor $\sim20-30$ in few cases, as also shown in their Fig. 6. This is linked to the fact that \cite{Kirkpatrick19} adopt a SMG-like CO SLED, mainly obtained from CO observations of sources with low AGN fraction, while \cite{Perna18} values reflect the different CO-SLED used in individual works, typically assuming a steeper CO-SLED \citep[e.g.][]{Carilli&Walter13}. Our results (Sect. \ref{sect:sled}) indicate that a steeper CO SLED is more appropriate for luminous QSOs. Indeed, we note that \cite{Kirkpatrick19} consider only CO detections, which may bias their analysis towards CO-luminous sources.


We compute the star formation efficiency in the host-galaxies of the WISSH QSOs, defined as \textit{(i)} the ratio of star formation rate per molecular gas mass SFE = SFR/$M_{\rm gas}$ in units of yr$^{-1}$ or, equivalently, as \textit{(ii)} the ratio between the far infrared luminosity and the CO(1$-$0) luminosity, that is SFE = $L_{\rm IR}/L^{\prime}_{\rm CO(1-0)}$, in units of  $L_\odot$/(K km s$^{-1}$ pc$^2$). In Figure \ref{fig:lcolfir.pdf} (right panel) we adopt the second definition and show the SFE evolution as a function of redshift of the WISSH QSOs,  which typically show high values of SFE in the range between 130 and 460 (hereafter in units of $L_{\odot}$/(K km s$^{-1}$ pc$^2$)), similar to the SFE>100 derived in most $z>1$ QSOs \citep[][]{Riechers11SFE,Krips12,Feruglio14}. In the case of \jquatt, we measure a moderate SFE$\sim65$, consistent with the average evolution of SFE in main sequence galaxies (solid line). Specifically, we show the expected SFE-$z$ evolutionary trend for a stellar mass $M_{*}=10^{11}$ \msun\ from \cite{Sargent14,Speagle14}, adapted to our empirical definition of SFE by using a CO luminosity to gas mass conversion factor $\alpha_{\rm CO}=3.6$ \citep{Daddi10,Accurso17}, hereafter in units of $M_{\odot}$/(K \kms\ pc$^2$), and the $L_{\rm IR}-$SFR relation from \cite{Kennicutt12}.
SMGs are characterised by a wide range of SFE $\in[20,600]$, intermediate between QSOs and MS galaxies, as also found by e.g. \cite{Magdis12,Yang17}. We note that SMGs with AGN activity from \cite{Bothwell13} are characterised by high SFE values, similar to the SFE of high-$z$ QSOs.
No evolution of the SFE with redshift is observed in the host-galaxy of QSOs at $1<z<6.5$, in agreement with previous compilation of AGN \citep{Feruglio14,Perna18}.

The standard approach in previous studies of molecular gas in QSOs was to assume a starburst-like conversion factor (that is $\alpha_{\rm CO}\sim0.8$) to derive gas masses \citep[e.g.][]{Perna18}. This is justified by the fact that many high-z QSOs have been found to reside in highly star-forming host-galaxies, with SFR of hundreds to thousands \msunyr\ \citep[e.g.][]{Solomon05,Carilli&Walter13,Combes18}, being characterised by compact disk sizes \citep[e.g.][]{Brusa18,Feruglio18} and high molecular gas excitation \citep[e.g.][see also Sect. \ref{sect:disc-sled}]{Gallerani14,Carniani19}. Several works have shown a dependence of $\alpha_{\rm CO}$ on the gas metallicity, and offset from MS, defined as $\Delta_{\rm MS}={\rm log}(sSFR/sSFR_{\rm MS})$, where $sSFR$ is the specific SFR for MS galaxies \citep[e.g.][]{Wolfire10,Genzel12,Accurso17}. Specifically, the conversion function by \cite{Accurso17} suggests a larger $\alpha_{\rm CO}\sim3.5$ for non-interacting galaxies with ${\rm log}(M_{*}/M_{\odot})\sim10.5-11.2$ and $\Delta_{\rm MS}\sim0.5$, typical of the high-z QSOs sample (Sect. \ref{sect:SFE}). However, in the case of mergers and galaxies with $\Delta_{\rm MS}>1.3$, the starburst-like conversion factor should be used \citep{Accurso17, Ivison11}. Accordingly, given the evidence that $\sim80\%$ of WISSH targets are in merging systems, we adopt $\alpha_{\rm CO}\sim0.8$ to infer the molecular gas mass. 
Table \ref{table:inferred} lists the resulting values of \mgas, which span about one order of magnitude from $\sim4.4\times10^9$ \msun\ to $\sim3.5\times10^{10}$ \msun. This implies a variety in the properties of the molecular gas reservoirs around WISSH QSOs, consistent with previous findings in $z\sim1-6$ QSOs \citep{Perna18}. 

Concerning \jquatt\ and \jdieci, for which we have no CO observations, an estimate of $M_{\rm gas}$ can be derived by adopting a gas-to-dust ratio (GDR) and dust masses computed in Sect. \ref{sect:sed}. In low$-z$ galaxies, a GDR $\sim100$ is typically observed \citep[e.g.][]{Draine07, Leroy11}. Studies of massive star forming galaxies and SMGs out to $z\sim3-5$, comparing CO-based gas masses and dust masses obtained from far-infrared continuum emission, found a possibly increasing GDR with redshift, with typical GDR $\sim120-250$ at $z\sim2-4$ \citep[e.g.][]{Saintonge13, Miettinen17}. In the case of WISSH QSOs, we derive gas-to-dust ratios in the range GDR $\in[100-300]$, with an average GDR$\simeq180$ (Table \ref{table:inferred}). By adopting the latter, we obtain log$(M_{\rm gas}/M{_\odot})\sim11.0$ and log$(M_{\rm gas}/M_{\odot})\sim9.3$ for \jquatt\ and \jdieci, respectively. An alternative approach to estimate $M_{\rm gas}$ consists in using the luminosity of the [CII] emission line as a tracer of the molecular gas mass. A correlation between $M_{\rm gas}$ and $L_{\rm [CII]}$ has indeed been reported for $z\sim2$ main-sequence galaxies \citep{Zanella18}:
\begin{equation}\label{eq:zanella}
    {\rm log} (M_{\rm gas}/M{_\odot}) = 1.02\times[{\rm log} (L_{\rm [CII]}/L{_\odot})+1.28]
\end{equation}
with a 0.3 dex scatter. By using Eq. \ref{eq:zanella} we obtain log$(M_{\rm gas}^{\rm [CII]}/M_{\odot})\sim11.1$ for \jquatt, which is almost identical to the dust-based gas mass, while the [CII]-based log$(M_{\rm gas}^{\rm [CII]}/M_{\odot})\sim9.9$ in \jdieci\ is a factor of $\sim4$ larger than the value inferred from $M_{\rm dust}$. However, because of its low ionisation potential, [CII] simultaneously traces the molecular, atomic, and ionized gas phase \citep[e.g.][]{Sargsyan12,Croxall17}. Therefore, depending on the relative contribution of the different gas phases, the total measured $L_{\rm [CII]}$ might be higher than the one arising from the molecular gas only: this would lead to overestimated [CII]-based \mgas. Alternatively, dust-masses might be underestimated because of the approximation in our SED fitting that dust can be modelled by a single dust temperature \citep[e.g.][]{Eales12,Aravena16}.

The ratio between the molecular gas mass and the star formation rate $\tau_{\rm dep} = M_{\rm gas}/{\rm SFR}$ represents the gas depletion timescale (which is equivalent to 1/SFE), corresponding to the amount of time required to exhaust all the reservoir of molecular gas at the current rate of star formation. In the case of WISSH QSOs, we divide \mgas\ for the AGN-corrected SFR computed in Sect. \ref{sect:sled}, finding $\tau_{\rm dep}\lesssim100$ Myr for all sources with CO-based measure of \mgas, with $\tau_{\rm dep}$ as small as $\sim20-30$ Myr in \jzdue, \jqnove\ and \jdsette. Concerning the two QSOs with dust-based \mgas, the $\tau_{\rm dep}\sim20$ Myr inferred for \jdieci\ suggests that the molecular gas reservoir will be converted into stars on a similar timescale compared to the other WISSH QSOs, while in the case of \jquatt, the large \mgas\ and modest SFR translate into a $\tau_{\rm dep}\sim360$ Myr. Our results indicate that all WISSH QSOs (except \jquatt) are characterised by small $\tau_{\rm dep}$, shorter by an average factor of $\sim30$ compared to the depletion timescale of MS galaxies of similar $M_{*}$ and $z$. This difference cannot be explained by the uncertainty on $\alpha_{\rm CO}$ only.

The high SFE and small $\tau_{\rm dep}$ observed in WISSH QSOs might be due to starburst activity in the host-galaxies of our targets \citep[e.g.][]{Aravena16,Yang17}. In fact, in the four WISSH targets for which we measure $M_{*}$ (see Sect. \ref{sect:fgas}) we find moderate offset from MS $\Delta_{\rm MS}\sim0.3-0.8$. Alternatively, it may be related to the impact of AGN feedback, which can reduce the molecular gas reservoir available for SF by removing and/or heating gas through the deposition of energy and momentum via AGN-driven outflows. The latter scenario will be further discussed in Sect. \ref{sect:fgas}.


\begin{figure*}
    \centering
    \includegraphics[width = 0.485\textwidth]{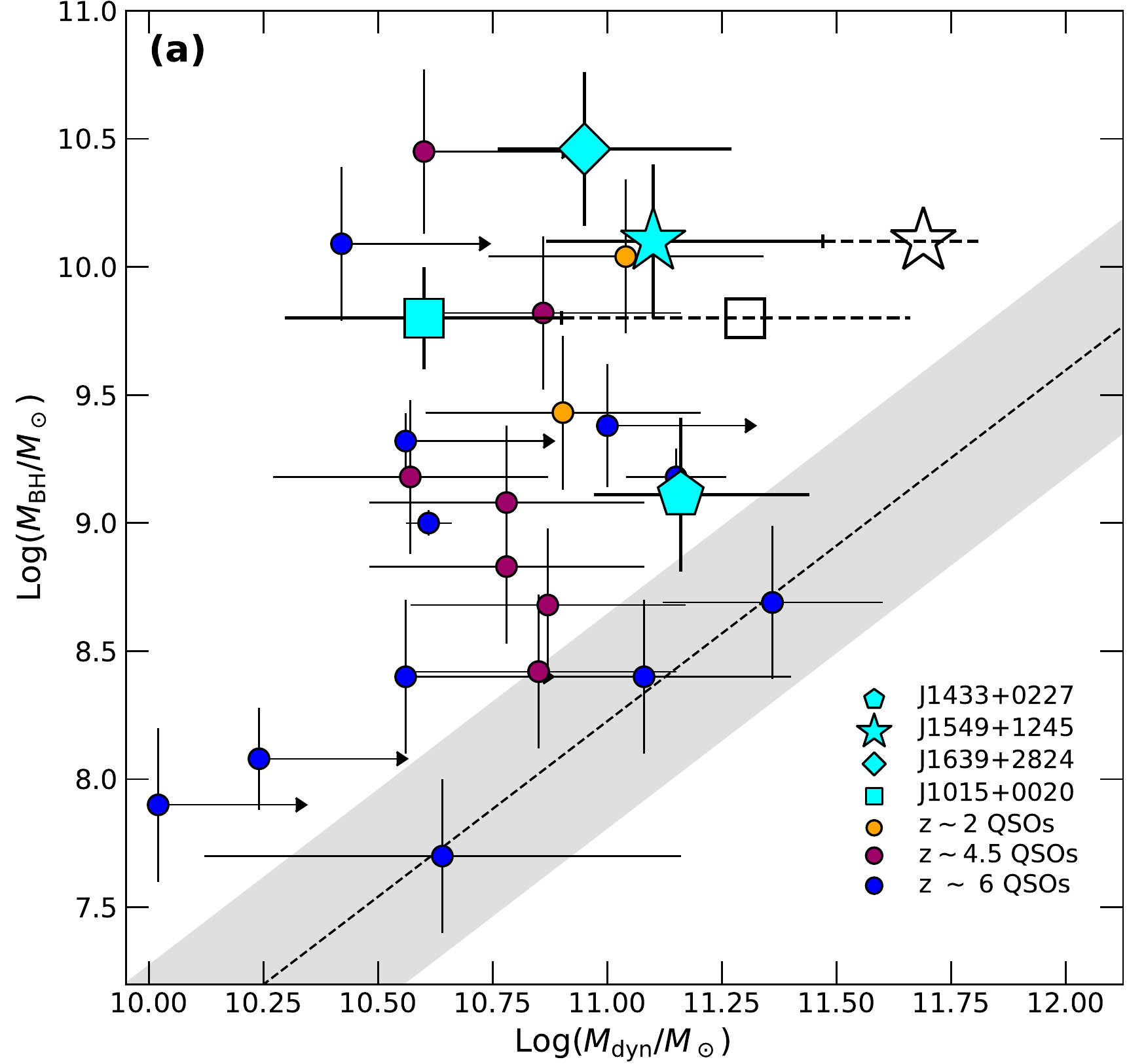}
    \includegraphics[width = 0.485\textwidth]{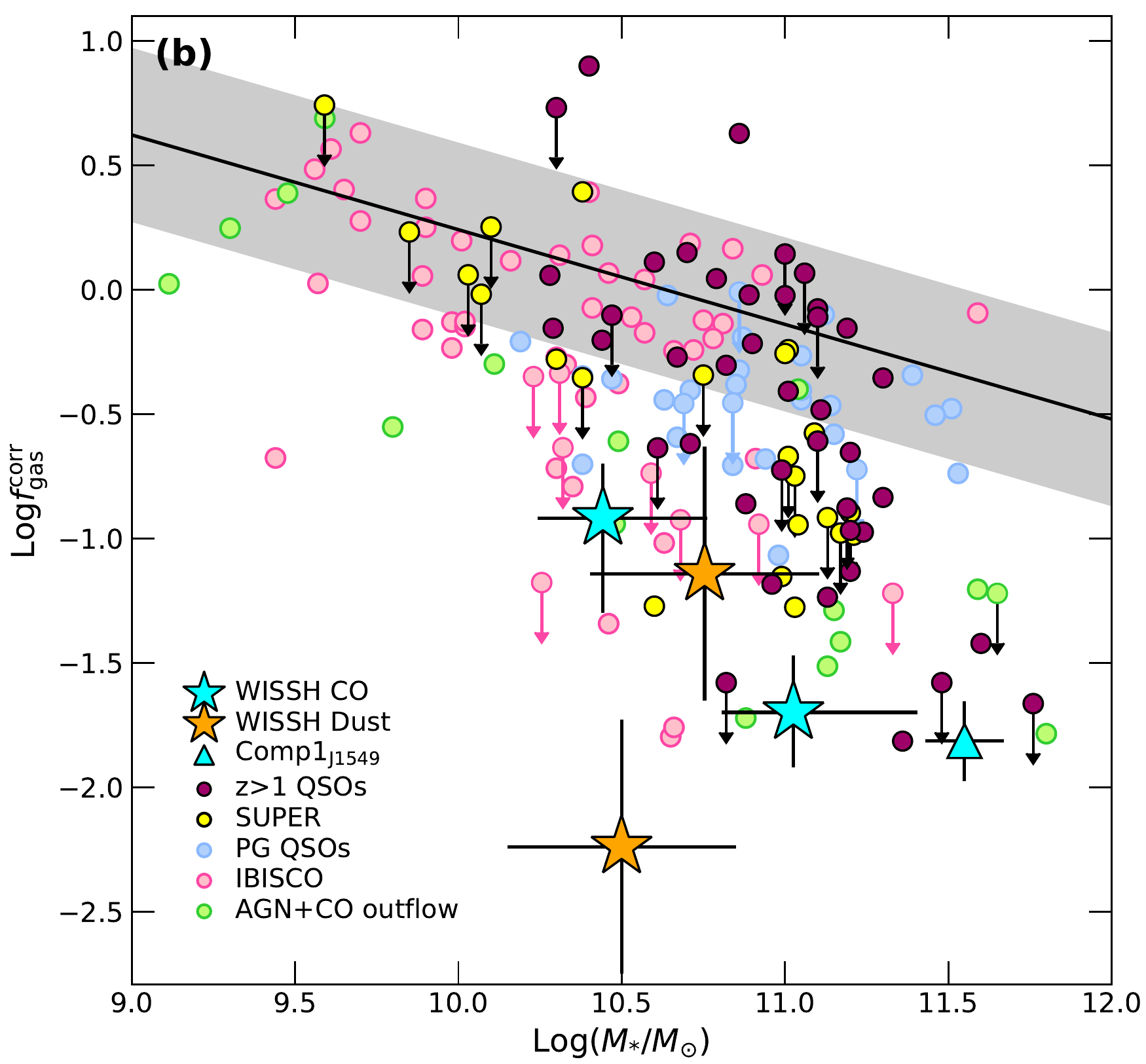}
    \caption{Panel (a) shows the black hole mass as a function of the dynamical mass of WISSH QSOs (cyan symbols), compared with $z\sim2-6$, luminous QSOs from literature, as indicated in the legend (see text for details). The total dynamical mass associated with the QSO + companion galaxies with measured $M_{\rm dyn}$ are also shown by the empty symbols. For WISSH QSOs and companions (as most literature sources), $M_{\rm dyn}$ values are based on the FWHM of the CO or [CII] emission lines. The $M_{\rm BH}-M_{\rm dyn}$ relation found for local galaxies by \cite{Jiang11} is also indicated by the dashed line, with the associated 0.4 dex intrinsic scatter (shaded region). Panel (b) displays the molecular gas fraction (corrected for the dependence on redshift and offset from main-sequence) as a function of the host-galaxy stellar mass for WISSH QSOs and a compilation of high-$z$ QSOs and local AGN (see text). WISSH QSOs with dust-based $M_{\rm gas}$ are indicated by orange stars. The average $f_{\rm gas}-M_{*}$ relation found for $z<4$ star-forming galaxies from the PHIBSS survey \citep{Tacconi18} is shown by the solid line. }
    \label{fig:mdyn-fgas}
\end{figure*}

\subsection{CO SLED}\label{sect:disc-sled}

The relative strengths of CO emission lines with different rotational quantum numbers $J$, that is the CO SLED (Sect. \ref{sect:sled}), provide key information on the contribution of the different gas phases to the total amount of molecular gas. Specifically, low-$J$ ($J\leq3$) CO lines trace the cold (excitation temperature $T_{\rm ex} \lesssim 100$ K), low-density (critical density $n_{\rm crit}\sim10^3-10^5$ cm$^{-3}$) phase, while high-$J$ ($J\geq5$) ones are associated with warmer and denser gas. Knowledge of the CO ladder is also necessary to convert measured high$-J$ CO transitions in CO(1$-$0) luminosity, CO(1$-$0) being the best known tracer of molecular gas mass ($M_{\rm gas}$) \citep[e.g.][]{Bolatto13}. It is relatively easy to probe this transition at cm wavelengths only in a minority of bright, high$-$z sources, as are our hyper-luminous QSOs. 

The CO SLED of high redshift QSOs is still poorly known, as it has been sampled only in few sources up to $J\gtrsim5$, including the CO(1$-$0) ground transition (Fig. \ref{fig:cosled}). However, the studied sources are known to be affected by strong gravitational lensing \citep{Downes95,Venturini03,Carilli03,Lewis02,Alloin07}, with possibly the exception of BR1202$-$0725 at $z\sim4.7$ \citep{Carilli02,Carniani13}. Differential magnification of the CO lines emitted by different regions may therefore bias the interpretation of their SLED. Both WISSH QSOs \jzdue\ and \jzotto, of which we investigate the CO SLED by combining CO(1$-$0) and CO(5$-$4) observations, are instead non-lensed. Specifically, SUBARU J,Ks and HST/WFC3 IR images of \jzdue\ \citep{Wang15} show no irregular surface brightness morphology, which may be indicative of multiple QSO images, down to an angular scale of $\sim0.2$ arcsec. Similarly, in the case of \jzotto, our inspection of {\it g} and {\it r} band images from SDSS DR12 archive showed no irregular morphology down to an angular scale of $\sim0.8$ arcsec. Moreover, in the case of BR1202$-$0725 and the QSO MG0751$+$2716 (also shown in Fig. \ref{fig:cosled}), the interpretation of the CO SLED might be further complicated by the presence of a radio jet \citep{Klamer04,Alloin07}. Similarly, ULAS J1234$+$0907 is characterised by strong AGN synchrotron emission \citep{Banerji18}, while our targets show no/little AGN contribution at radio wavelengths (Sect. \ref{sect:sed}).

The CO SLED of WISSH QSOs calculated in Sect. \ref{sect:sled} shows a steep rise up to $J=5$. Specifically, the CO SLED of \jzdue\ is consistent with previous measurements of CO excitation in high-z QSOs \citep{Carilli&Walter13,Casey14}, while in the case of \jzotto\ the observed CO(5$-$4)/CO(1$-$0) ratio is even higher than that expected in APM08279$+$5255, that is the QSO with the steepest CO SLED observed so far \citep{Weiss07}. We note that a similarly high CO(5$-$4)/CO(1$-$0) ratio to that of \jzotto\ has been recently reported for a weakly lensed, Type 1 QSO at $z\sim2.8$ by \cite{Fogasy20}.
The high excitation of the molecular gas in our hyper-luminous targets is likely linked to AGN-related heating agents, such as X-ray photons or shocks, which influence molecular gas physics and chemistry in X-ray dominated regions  \citep[XDRs, ][]{Maloney96}. In these regions, molecular gas can be heated above 100 K \citep{VanDerWerf10,Meijerink13, Mingozzi18,Vallini19,Li20} and the high$-J$ CO rotational energy levels become more populated with respect to photo-dissociation regions \citep[PDRs,][]{Hollenbach99}, dominated by UV photons from young stars, which mostly contribute to the CO SLED of non active, star forming galaxies \citep[e.g.][]{Narayanan14,Greve14}. The steep CO SLED observed in Comp$_{\rm J0209}$ suggests that the AGN radiative output might be able to affect cold gas excitation also in nearby sources.

The limited sampling of the CO ladder in \jzdue\ and \jzotto\ as well as in their companion galaxies does not allow us to quantify the different contributions to CO excitation. Further observations of high$-J$ CO rotational transitions, that is $J\sim6-10$, close to the peak of the CO SLED \citep{Mashian15,Li20}, will be important to probe the physical properties of the molecular gas and the main excitation source in these luminous targets.

\subsection{Dynamical masses}\label{sect:mdyn}

\subsubsection{Size of CO and [CII] emission}
In Sect. \ref{sect:linemaps} and Sect. \ref{sect:kine} we investigated the CO and [CII] spatial extent and kinematics of the WISSH QSOs \jquatt, \jqnove\ and \jsedici, in which line emission is spatially resolved by the ALMA observations. Specifically, we find the cold gas reservoir in \jquatt\, as traced by [CII] emission, to be distributed in a rotating disk with size $D=4.4\pm0.3$ kpc. The latter was computed by multiplying the major axis derived in Sect. \ref{sect:linemaps} by a factor of 1.5 \citep[e.g.][]{Wang13,Venemans16}. This value is in agreement with typical sizes ($2-5$ kpc) of [CII] emission measured in luminous, high-z QSOs at $z\sim4.5-5$ \citep{Trakhtenbrot17, Pensabene20, Nguyen20}, and a factor of three larger than the [CII] disk size of the WISSH QSO \jdieci\ \citep{Bischetti18}.

In the case of \jqnove, the CO(4$-$3) emission is peculiarly extended over $D = 12.5\pm2.0$ kpc, as also observed in the companion galaxy Comp1$_{J1549}$, which is characterised by an even larger CO(4$-$3) size $D=15.4\pm1.9$ kpc for the molecular gas reservoir. The large size measured in \jqnove\ differs from the majority of previous observations targeting mid$-J$ CO rotational transitions in QSO host-galaxies, finding the bulk of molecular gas located in compact regions of few kpc size \citep[e.g.][]{Fan18,Bischetti19,DAmato2020}. On the other hand, CO(3$-$2) emitting regions with $D>10$ kpc have been measured in hyper-luminous, reddened QSOs at $z\sim2.5$ by \cite{Banerji17}, with comparable \lbol. Given that several SMGs show evidence for similarly extended reservoirs but in lower excitation molecular gas \citep[e.g.][]{Ivison11, Riechers11size}, this may suggest that the huge QSO radiative output in our targets is able to affect CO excitation out to tens of kpc scale and in nearby galaxies, in agreement with our findings for the CO SLED of \jzdue\ and its companion (Sect. \ref{sect:sled}).
Concerning \jsedici, we instead measure a compact size for the CO(4$-$3) emission $D=1.7\pm0.2$ kpc, which is a factor of two larger than the CO size measured for this QSO by \cite{Schramm19} from the same observation. We note that the detection of gas on larger scales might be hampered by the very high native angular resolution of the ALMA observations (Sect. \ref{sect:data}).

\subsubsection{Dynamical vs SMBH mass}
To calculate the dynamical mass of \jquatt, we {\it(a)} adopt the virial relation $M_{\rm dyn}=D_v v_{\rm rot}^2/2G=5.6_{-1.9}^{+5.1}\times10^{10}$ \msun, where $v_{\rm rot}$ is provided by the BAROLO model of the high-resolution ALMA datacube (Sect. \ref{sect:kine}). As the BAROLO model is limited to [CII] emission from the central region with (deconvolved) size of  $0.31\pm0.03$ arcsec, the virial estimate corresponds to the dynamical mass within the inner $D_v=3.5\pm0.3$ kpc; {\it(b)} use the FWHM of the [CII] emission line as a proxy of the circular velocity, according to the relation $M_{\rm dyn}=1.16\times10^5\times(0.75\times{\rm FWHM}/{\rm sin}i)^2\times~D=1.4_{-0.5}^{+1.3}\times10^{11}$ \msun\ \citep[e.g.][]{Wang13,Venemans16}.  Similarly, by using {\it(b)} we find $M_{\rm dyn}=1.2_{-0.5}^{+1.7}\times10^{11}$ \msun\ for \jqnove, and a larger mass by a factor of three $M_{\rm dyn}=3.6_{-0.4}^{+1.1}\times10^{11}$ \msun\ for its companion galaxy Comp1$_{\rm J1549}$, supporting an interpretation of the QSO-companion interaction as major merger. The measured dynamical mass of \jsedici\ is $M_{\rm dyn}=8.9_{-3.2}^{+9.7}\times10^{10}$ \msun, which is about one order of magnitude larger than the value in \cite{Schramm19} which, however, reported their estimate of $M_{\rm dyn}$ for a rotating disk to be smaller than the black hole mass for this QSO. We point out that in the case of marginally resolved sources such as our targets, detected with moderate significance, inclination and, in turn, $M_{\rm dyn}$ estimates can be significantly altered by non-circular beam shapes. Also, non-rotating ISM components may be missed given the angular resolution of our observations. Moreover, our $M_{\rm dyn}$ are based on a single emission line whose kinematics may represent only the inner regions of the galaxy \citep[e.g.][]{DeBlok14,Lupi19}. Further discussion about these issues can be found in \cite{Trakhtenbrot17} and references therein. Finally, we note that assuming WISSH systems to be dispersion-dominated would result in smaller M$_{\rm dyn}$ by a factor of $3-4$ \citep[e.g.][]{Decarli18}. 

Both \jqnove\ and \jsedici\ benefit from single-epoch measurements of the SMBH mass ($M_{\rm BH}$) based on the H$\beta$ $\lambda4861$ $\AA$ emission line which, being mostly dominated by virial motions, is the best estimator of $M_{\rm BH}$ \citep{Denney12,Marziani12,Vietri18} for QSOs up to $z\sim3.8$. Specifically, \cite{Bischetti17} and \cite{Schramm19} found very large ${\rm log}(M_{\rm BH}/M{_\odot})\simeq10.1-10.4$. For the higher redshift QSO \jquatt, a ${\rm log}(M_{\rm BH}/M{_\odot})\simeq9.11$ measured from the MgII $\lambda2800$ $\AA$ emission line was reported by \cite{Trakhtenbrot11}.

Figure \ref{fig:mdyn-fgas}a shows the location of the WISSH QSOs in the $M_{\rm BH}-M_{\rm dyn}$ plane (cyan stars), including \jdieci\ analysed in \cite{Bischetti18}. QSOs from literature with measure of $M_{\rm dyn}$ based on CO or [CII] emission lines and single epoch estimate of $M_{\rm BH}$ are also displayed. We include dusty-obscured QSOs at $z\sim2$ from \cite{Banerji15,Banerji17,Bongiorno14,Brusa18} and luminous $z\sim4.5-6$ QSOs from the works of \cite{Venemans16,Venemans17,Willott13,Willott15,Willott17,Kimball15,Trakhtenbrot17,Feruglio18,Mortlock11,DeRosa14,Kashikawa15}. The $M_{\rm BH}-M_{\rm dyn}$ relation, derived from local galaxies in a wide range of $M_{\rm dyn}\sim10^9-10^{12}$\msun\ by \cite{Jiang11}, is also shown for comparison. All WISSH QSOs but \jquatt\ are characterised by a ratio between dynamical mass and SMBH mass $M_{\rm dyn}/M_{\rm BH}\sim3-10$, among the smallest observed so far. According to the local relation, the typical $M_{\rm dyn}/M_{\rm BH}$ ratio should be $\sim600$ which, given the very massive black holes hosted by our targets, should translate into host-galaxy dynamical masses $\gtrsim10^{12}$ \msun. Such large values of $M_{\rm dyn}$ suggest that that hyper-luminous QSOs are the likely placeholders of high-density regions were local giant galaxies have been assembled \citep[see also][]{Jones17,Diaz-Santos18}. 
We also note that small $M_{\rm dyn}/M_{\rm BH}$ ratios are expected in the early growth phases ($z\gtrsim3$) of 
very massive BHs, for a scenario in which black-hole accretion is triggered by galaxy interactions \citep[e.g.][]{Lamastra10,Menci14}.

In the case of \jqnove, the empty star in Fig. \ref{fig:mdyn-fgas}a represents the total dynamical mass ${\rm log}(M_{\rm dyn}/M{_\odot})\sim11.68$ associated with the QSO plus Comp1$_{\rm J1549}$ system, under the likely hypothesis that these two sources will merge (see Sect. \ref{sect:linemaps}) and build up the mass of the QSO host-galaxy. A similar result was reported for \jdieci\ (whose mass plus that of its companion galaxy is also reported in Fig. \ref{fig:mdyn-fgas}a as the empty square) by \cite{Bischetti18}, who found this QSO to be associated with multiple nearby (< 16 kpc) companion galaxies, for a total a ${\rm log}(M_{\rm dyn}/M{_\odot})>11$ already in place at $z\sim4.4$. Given the widespread presence of bright companion galaxies in the surroundings of WISSH QSOs, most of which are located at close projected distance ($\sim6-30$ kpc), we expect them to significantly contribute to the final mass of the QSO host-galaxies and consequently move the WISSH points closer to the local relation.


\subsection{Gas fraction and feedback from QSO winds}\label{sect:fgas}

To probe the amount of gas available for star-formation activity in the WISSH QSOs, we compute the molecular gas fraction, defined as $f_{\rm gas}=M_{\rm gas}/M_{*}$. Our targets being unobscured QSOs, a derivation of the host-galaxy stellar mass from SED fitting is particularly challenging, given that the nuclear emission outshines the galaxy at  wavelengths shorter than few tens of microns (Sect. \ref{sect:sed}). An alternative approach is to infer $M_{*}$ via the dynamical mass, according to the relation $M_{*}=M_{\rm dyn}-M_{\rm gas}-M_{\rm BH}$ \citep[e.g.][]{Venemans17b, Nguyen20}, under the assumption that dark matter does not significantly contribute to the central mass of the host-galaxies \citep[e.g.][]{Wuyts16}. By using the latter method, we could therefore measure $M_{*}$ for four of our targets and we find ${\rm log}(M_{*}/M{_\odot})\sim10.4-11.0$ and a wide range of $f_{\rm gas}\sim0.04-1.6$ for our targets, similarly to what has been found in other high$-z$ QSOs \citep[e.g.][]{Venemans17b,Banerji18}. However, we caution that large uncertainties affect these measurements, the main of which are related to the uncertainty on the galaxy inclination and to the assumption of an $\alpha_{\rm CO}$ (Sect. \ref{sect:SFE}). This prevents us from putting tight constraints on $M_{*}$ and, in turn, on $f_{\rm gas}$. 
Previous studies of molecular gas content in star-forming galaxies out to $z\sim4$  \citep[e.g.][]{Genzel15,Tacconi18} highlighted a dependence of $f_{\rm gas}$ on redshift and offset from the star-forming galaxies MS, $\Delta_{\rm MS}$. To ensure a meaningful comparison of the molecular gas fraction of WISSH QSOs with other samples at different $z$ and $\Delta_{\rm MS}$, we thus correct $f_{\rm gas}$ for these trends. Specifically, we use the functions $f_2(z)$ and $g_2(sSFR/sSFR_{\rm MS})$ by \cite{Genzel15,Tacconi18}, in which we parametrise $sSFR_{\rm MS}$ according to \cite{Whitaker12}. 

Figure \ref{fig:mdyn-fgas}b shows $f_{\rm gas}^{\rm corr}$ (corrected for trends with $z$ and $\Delta_{\rm MS}$) of WISSH QSOs as a function of M$_{*}$, compared with QSO samples as in Sect. \ref{sect:SFE}. We also include low-redshift AGN samples, such as {\it (i)} Palomar Green (PG) QSOs with CO(2$-$1) based molecular gas masses \citep{Shangguan20}; {\it (ii)} X-ray selected {\it INTEGRAL}/IBIS AGN, characterised by log$(L_{\rm Bol}/{\rm erg\ s^{-1}})\sim43.5-45.6$, with CO(1$-$0) and CO(2$-$1) measurements from the IBISCO survey (Feruglio et al. 2020 in prep.). We consider {\it (iii)} active galaxies with evidence of AGN-driven molecular outflows from  \cite{Fiore17, Brusa18, Fluetsch19,Herrera-Camus19}. For all these sources, molecular gas masses have been homogeneously computed as in Sect. \ref{sect:SFE}. The average $f_{\rm gas}^{\rm corr}-M_{*}$ relation derived from  $\sim1400$ star forming galaxies at $z<4$ from the IRAM Plateau de Bure HIgh-z Blue Sequence Survey \citep[PHIBBS,][]{Tacconi18} is indicated by the solid line. 

We find that the molecular gas fraction of WISSH QSOs is
systematically smaller by a factor of $\sim10-100$ than that of star-forming galaxies with the same stellar mass. In general, AGN with ${\rm log}(M_{*}/M{_\odot})<10.25$ exhibit similar $f_{\rm gas}^{\rm corr}$ to star-forming galaxies, while at larger stellar masses the number of sources below the relation significantly increases, both in high-z QSOs and in $z\sim0$ AGN. We note that part of the offset of WISSH QSOs (and other AGN in starbursting or merging systems, see Sect. \ref{sect:SFE}) might be due to the adopted $\alpha_{\rm CO}=0.8$, while the molecular gas fraction in PHIBBS galaxies is based on a complex conversion function, which takes into account metallicity and offset from MS \citep[][]{Genzel15,Accurso17}. This may account for a factor of $\sim4$ in $f_{\rm gas}^{\rm corr}$ (error bar in Fig. \ref{fig:mdyn-fgas}b), given the $M_{*}$ of our targets. However, even taking into account the uncertainty on $\alpha_{\rm CO}$, WISSH QSOs and $\sim50\%$ of points at ${\rm log}(M_{*}/M{_\odot})\gtrsim10.5$ fall short of the \cite{Tacconi18} average $f_{\rm gas}^{\rm corr}$.

This result is in agreement with previous findings by \cite{Brusa15,Perna18}, who reported a lower molecular gas fraction in obscured QSOs compared to MS galaxies and proposed AGN feedback as responsible for depleting the host-galaxy of the molecular gas content via outflows, during the transition, "blowout" phase from starburst galaxy to unobscured QSO \citep[e.g.][]{Menci08,Hopkins08}. Moreover, \cite{Brusa18} reported a low gas fraction in the $z\sim1.5$ luminous, obscured QSO XID2028, in which an outflow has been detected in both the molecular and ionised gas phases \citep[see also][]{Cresci15} 
Similarly, \cite{Fiore17} reported a lower $f_{\rm gas}^{\rm corr}$ than MS galaxies in a collection of local AGN with evidence of molecular outflows, the gap increasing at large $M_{*}$. \cite{Carniani17} targeted $L_{\rm Bol}\sim10^{47}$ \ergs\ QSOs at $z\sim2.4$ showing strong ionised winds, as traced by the [OIII] $\lambda5007$ $\AA$ emission line, and reported a significantly reduced gas reservoir compared to main-sequence galaxies at the same redshift. On the other hand, \cite{Herrera-Camus19} found a $f_{\rm gas}^{\rm corr}$ consistent with the \cite{Tacconi18} relation in a $z\sim2.4$ QSO hosted by a massive ($M_{*}\sim10^{11}$ \msun), MS galaxy, with multi-phase outflows.

It is worth noting that all WISSH QSOs analysed in this work show evidence of ionised outflows, from nuclear to circum-galactic scales, as
traced by large blue-shifts of their emission lines in the rest-frame UV and optical spectrum with respect to the systemic QSO redshift, as traced by CO or [CII] ($z_{\rm cold}$ in Table \ref{table:co-properties}). Specifically, by analysing SDSS DR12 spectra \citep{Alam15} we measured the wavelength corresponding to the peak of the CIV $\lambda1549$ $\AA$ emission line  ($z_{\rm SDSS}$ in Table \ref{table:sample}), and found large velocity blue-shifts in the range $v\in[1200-5600]$ \kms with respect to the systemic redshift.
Similar results were found by \cite{Vietri18} who analysed the velocity shift between CIV and H$\beta$ for 18 WISSH QSOs and revealed powerful, broad-line region winds with associated kinetic power $\dot{E}_{\rm kin}\sim10^{43}-10^{44}$ \ergs. The presence of strong nuclear winds in $\sim25\%$ of the total WISSH sample was also reported in \cite{Bruni19}, who found high velocity BAL features blue-wards of CIV and SiIV $\lambda1400$ $\AA$ emission lines. Three WISSH among the targets of this work are BAL QSOs, namely \jqnove, \jqcinque, and \jqotto, which shows an ultra-fast BAL outflow with velocity $v\in38000-47000$ \kms.
Indeed, \cite{Bischetti17} reported among the most powerful [OIII] outflows observed so far in five WISSH QSOs (including \jqnove), with $\dot{E}_{\rm kin}$ up to few percent of $L_{\rm Bol}$ and extending on kpc scale, while \cite{Travascio20} discovered a Ly$\alpha$ $\lambda 1216$ $\AA$ outflow  propagating up to $\sim30$ kpc in the circum-galactic medium of \jqotto. Feedback associated with QSO activity, responsible for heating and/or removing molecular gas from the host-galaxy via outflows, is a possible scenario for the low molecular gas fractions. 

We searched for the presence of cold gas outflows as traced by broad/asymmetric wings in the CO or [CII] spectra of the WISSH QSOs presented in this work. We found no detection of high-velocity cold gas. However, the limited sensitivity of our observations (Table \ref{table:sample}) prevents us from putting meaningful upper limits on the luminosity of cold gas outflows in our targets. Indeed, molecular and neutral outflows observed in luminous QSOs typically correspond to $\sim1/60$ to $\sim1/20$ of the peak of CO and [CII] emission line profiles, respectively \citep[e.g.][]{Feruglio17,Bischetti19,Cicone15,Bischetti19b}.
Therefore, deeper observations in the (sub-)~millimetre band are required to detect the counterpart in the cold gas phase of the powerful ionised outflows in WISSH QSOs.

\section{Conclusions}\label{sect:conclusion}

We report on ALMA, NOEMA and JVLA observations of the far-infrared continuum, CO and [CII] line emission in a sample of eight hyper-luminous QSOs from the WISSH sample at $z\sim2.4-4.7$. These data enable us to perform the first systematic study of cold gas properties in hyper-luminous QSOs at Cosmic noon. Our main findings can be summarised as follows:
\begin{itemize}
    \item[--] We detect CO(4$-$3), CO(5$-$4) rotational  or [CII] emission in the host-galaxies of our targets ($100 \%$ detection rate). In the case of \jzdue\ we also detect CO(1$-$0). We find CO emission to be highly excited, as suggested by the CO SLEDs of \jzdue\ and \jzotto, among the steepest observed in high-z QSOs so far. Far-infrared and/or mm continuum is detected in $\sim75\%$ of our targets.
    \item[--] For spatially resolved sources, we observe a variety of sizes for the molecular gas reservoirs, in the range $\sim1.7-10$ kpc. In \jquatt, our dynamical modelling of the velocity gradient observed in the [CII] emission line indicates the presence of a fast rotating disk, in which gas motions not associated with rotation are also present. Similarly, disturbed kinematics is observed in \jqnove\ and \jsedici. 
    \item[--] Our hyper-luminous QSOs are preferentially located in high-density regions, given the widespread presence of one or more line emitters around $\sim80 \%$ of the QSOs. These sub-millimetre companions show bright CO or [CII] emission, in most cases comparable to emission from cold gas in the QSO host-galaxy, and are located at projected distances of $\sim6-130$ kpc.
    \item[--] We find that the majority of WISSH QSOs exhibit a lower $L^{\prime}_{\rm CO}$ (by a factor of $\sim4$) than $z\sim0-3$ MS galaxies with same $L_{\rm IR}$, after correcting for the AGN contribution to infrared wavelengths. This translates into SFE $>100$ $L_\odot$/(K km s$^{-1}$ pc$^2$), implying that the observed SFRs $\in[165-1095]$ \msunyr\ are able to convert the molecular gas reservoir into stars on a timescale of $\lesssim50$ Myr, shorter by an average factor of $\sim30$ compared to the depletion timescale of MS galaxies of similar $M_*$ and $z$.  
    \item[--]  All WISSH QSOs  but \jquatt\ are  characterised  by extremely small $M_{\rm dyn}/M_{\rm BH}$ ratios of $\sim3-10$, which are about two orders of magnitude offset from local relations. We find hyper-luminous QSOs to pinpoint the high-density sites where giant galaxies assemble, with significant contribution of mergers to the host-galaxy mass. 
    \item[--] We also infer the molecular gas fraction in the host-galaxies of four WISSH QSOs, finding lower values by a factor of $\sim10-100$ once compared to star forming galaxies with same $M_{*}$. This is likely linked to the widespread evidence of AGN-driven outflows in our targets having the effect of heating/depleting the gas reservoir in the galaxy.

\end{itemize}

The ALMA, JVLA and NOEMA observations analysed in this work provide a wealth of information about the properties of cold ISM and environment for a homogeneous and statistically fair sample of $L_{\rm Bol}>10^{47}$ \ergs\ Type 1 QSOs shining at the peak epoch of QSO activity and massive galaxy assembly.
We find these hyper-luminous objects to be caught in an evolutionary phase of concurrent intense growth of both SMBH and host-galaxy. Such growth is likely being regulated by AGN-feedback and will use up the molecular gas reservoir in few tens of Myr, supporting a scenario in which hyper-luminous QSOs are the progenitors of  "red  \&  dead" giant ellipticals. Given the richness of companion galaxies, we expect these systems to further evolve towards the local $M_{\rm BH}-M_{\rm dyn}$ relation via mergers.

The results presented here thus represent a critical step towards a better understanding of the SMBH-galaxy co-evolution for the objects at the extreme end of the mass function.
Follow-up observations with higher angular resolution of these QSOs are needed to optimally map the cold gas kinematics in their hosts and in the companion galaxies, and better constrain the total mass and the merger state of these systems. Such observations may also be able to detect the cold gas counterparts of the ionised outflows revealed in all WISSH QSOs, assessing their multi-phase nature and global energetics.

\begin{acknowledgements}
This paper makes use of the following ALMA data: ADS/JAO.ALMA\#2013.1.00417.S, ADS/JAO.ALMA\#2019.1.01070.S, ADS/JAO.ALMA\#2015.1.01602.S, ADS/JAO.ALMA\#2016.1.00718.S, ADS/JAO.ALMA\#2016.1.01515.S, ADS/JAO.ALMA\#2018.1.01806.S. ALMA is a partnership of ESO (representing its member states), NSF (USA) and NINS (Japan), together with NRC (Canada), MOST and ASIAA (Taiwan), and KASI (Republic of Korea), in cooperation with the Republic of Chile. The Joint ALMA Observatory is operated by ESO, AUI/NRAO and NAOJ. This work is based on observations carried out under project number S17BW and W17DT with the IRAM NOEMA Interferometer. IRAM is supported by INSU/CNRS (France), MPG (Germany) and IGN (Spain). We are grateful to the anonymous referee for useful feedback which helped us to improve the paper. We thank Dr. Y. Wang for providing us with SUBARU images of \jzdue\ and Dr. P. Santini for useful discussion about SFE in high$-z$ galaxies. 
MBi, MBr, CF, FF, AM, EP acknowledge support from PRIN MIUR project "Black Hole winds and the Baryon Life Cycle of Galaxies: the stone-guest at the galaxy evolution supper", contract \#2017PH3WAT. MBi, CF and FF acknowledge support from INAF under PRIN SKA/CTA FORECaST and PRIN MAINSTREAM 2018 "Black hole winds and the baryon cycle".
EP, LZ and MBi acknowledge financial support under ASI-INAF contract 2017-14-H.0. 
GB acknowledges financial support under the INTEGRAL ASI-INAF agreement 2019-35-HH.0
MPT acknowledges financial support from the State Agency for Research of the Spanish MCIU through the Center of
Excellence Severo Ochoa award to the Instituto de Astrof\'isica de Andaluc\'ia (SEV-2017-0709) and through grant PGC2018-098915-B-C21 (MCI/AEI/FEDER, UE). RM acknowledges ERC Advanced Grant 695671 “QUENCH” and support by the Science and Technology Facilities Council (STFC). GVe acknowledges support from CONICYT Basal-CATA AFB-170002 and FONDECYT Postdoctorado 3200802 grants. CC acknowledges support from the Royal Society.
MP is supported by the Programa Atracci\'on de Talento de la Comunidad de Madrid via grant 2018-T2/TIC-11715. GVi acknowledges financial support from Premiale 2015 MITiC (PI B. Garilli).

\end{acknowledgements}


\bibliography{biblio.bib}

\newpage
\onecolumn
\section*{Appendix: broad-band UV-to-FIR SEDs of WISSH QSOs}
\begin{figure*}[htb]
	\centering
	\caption{Rest-frame SED of the WISSH QSOs considered in this work. In each panel, black symbols indicate the photometric points considered in our modelling. Black circles identify detections while arrows represent $3\sigma$ upper limits (see Table. \ref{table:sample}). Photometric points at $\lambda<1216 \AA$ are not included in the fits due to Ly$\alpha$ absorption (grey circles). Black curve represents the total best fit model, while blue and orange curves refer to the accretion disk plus torus and cold dust emission, respectively. In the case of \jqcinque, the green curve represents the best-fit template reproducing the near-IR excess \citep[for details, see][]{Duras17}.}
	
    \includegraphics[width=1\textwidth]{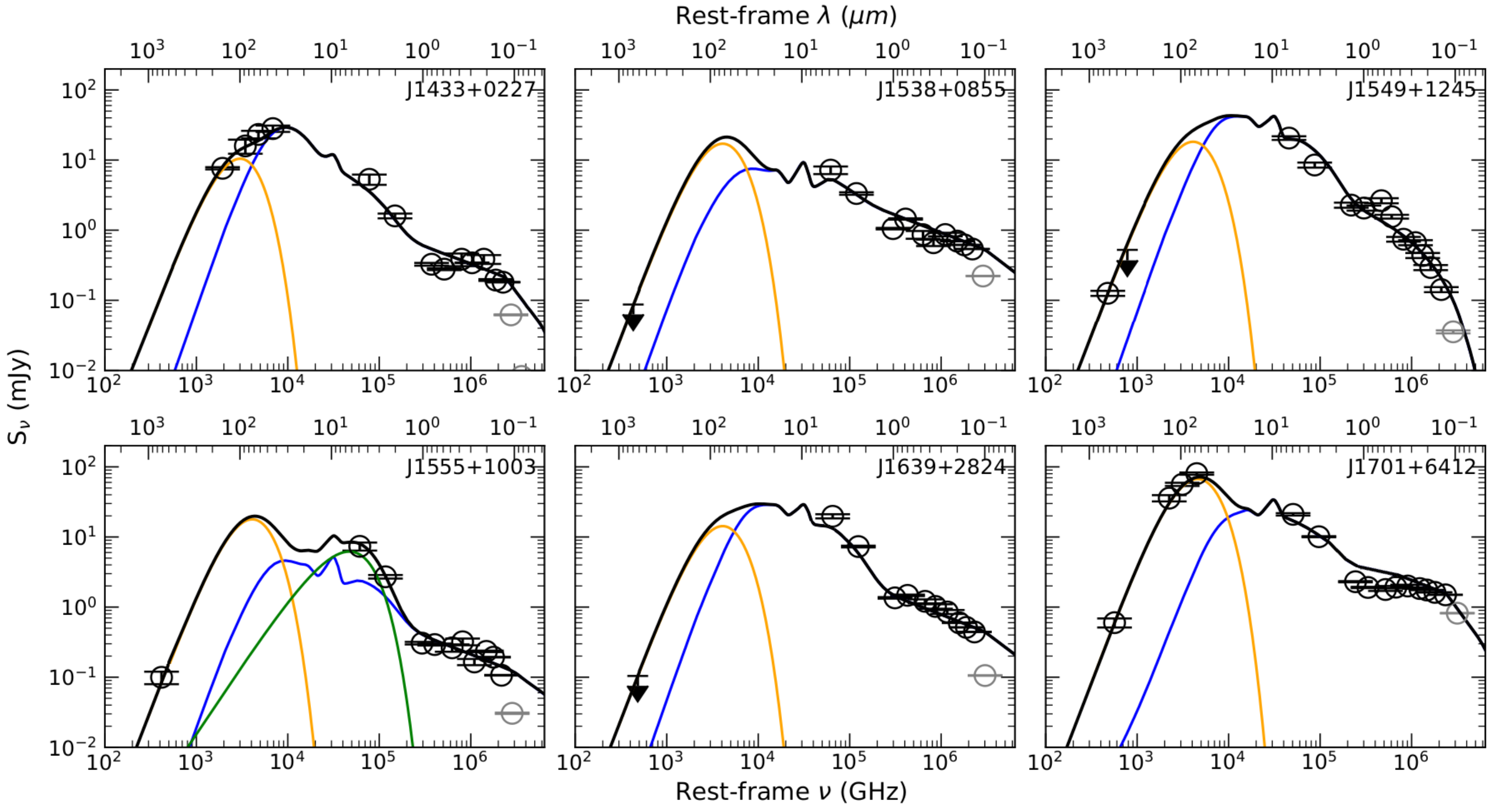}
	\label{fig:sed}
\end{figure*}
\end{document}